\documentstyle[epsf]{mn}

\title[Field ellipticals at $z\approx 0.3$ II]{The properties of field
elliptical galaxies at intermediate redshift.  II: photometry and
spectroscopy of an HST selected sample}

\author[T.~Treu et al.]{T.~Treu,$^{1,2,4}$ M.~Stiavelli,$^{2}$
 P.~M{\o}ller,$^{3}$ S.~Casertano$^{2}$ and G.~Bertin$^1$\\
$^1$ Scuola Normale Superiore, P.za dei Cavalieri 7, I-56126, Pisa,
Italy\\
$^2$ Space Telescope Science Institute, 3700 San Martin Dr.,
Baltimore, MD 21218, USA\\
$^3$ ESO, Karl-Schwarzschild Str. 2, D85748, Garching bei M\"unchen,
Germany\\
$^4$ present address: California Institute of Technology, Astronomy 105-24, Pasadena, CA 91125, USA; e-mail tt@astro.caltech.edu\\ 
}

\newcommand{\tx}[1]{\textrm{#1}}
\newcommand{\pc}[1]{\protect\citename{#1}}

\newcommand{\kms}{km~$\tx{s}^{-1}$}

\newcommand{\dv}{$r^{1/4}\,$}
\newcommand{\sgh}{$\sigma_{\tx{ghff}}$\,}

\newcommand{\zgh}{$z_{\tx{ghff}}$\,}

\newcommand{\Io}{I$_{814}$}
\newcommand{\Vs}{V$_{606}$}
\begin{document}
\maketitle

\begin{abstract}
A sample of field early-type galaxies (E/S0) at intermediate redshift
($z\sim0.1-0.6$) is selected, based on morphology and colours from
HST-WFPC2 parallel images.  Photometric structural parameters
(effective radius $R_{\tx{e}}$ and effective surface brightness
$SB_{\tx{e}}$) are derived through the F606W and F814W filters, using
luminosity profile fitting and two-dimensional fitting techniques. The
combined parameter that enters the Fundamental Plane ($\log
R_{\tx{e}}-\beta SB_{\tx{e}}$, with $\beta\approx0.32$) is shown to
suffer from significantly smaller uncertainties (r.m.s. 0.03) than the
individual structural parameters (e.g. $\sim 15$ per cent r.m.s. on the
effective radius).

High signal-to-noise intermediate resolution spectra, taken at the
ESO-3.6m, yield redshifts for 35 galaxies and central velocity
dispersions for 22 galaxies. Central velocity dispersions are derived
using a library of stellar templates covering a wide range of spectral
types, in order to study the effects of templates mismatches. The
average random error on the central velocity dispersion is found to be
8 per cent and the average systematic error due to template mismatch
is found to be 5 per cent. The errors on velocity dispersion
measurement and the effects of template mismatches are studied by
means of extensive Montecarlo simulations.  In addition, we
investigate whether the determination of the velocity dispersion is
sensitive to the spectral range used, finding that the value of
velocity dispersion is unchanged when the spectral regions that
include the absorption features Ca HK and NaD are masked out during
the fit.

\end{abstract}

\section{Introduction}
\label{sec:intro}

This paper is the second of a series devoted to the investigation of
the internal and structural properties of field early-type galaxies
(E/S0) at intermediate redshift ($0.1<z<1$), and of the properties of
their stellar populations. In particular, the aim of this series of
papers is to investigate how empirical scaling laws such as the
Fundamental Plane (FP; Djorgovski \& Davis 1987; Dressler et al.\
1987) and the Kormendy (1977) relation evolve with redshift.  These
studies are complemented with the measurement of metallicity and age
of stellar populations by means of the Lick/IDS absorption lines
indeces (Trager et al.\ 1998).

For this project we have collected high signal-to-noise (S/N) spectra
at intermediate resolution (resolving power
$\lambda/\Delta\lambda\sim1000$, where $\Delta \lambda$ is the full
width half maximum resolution) for a well-defined sample of field
early-type galaxies in the redshift range $z\approx 0.1-0.6$. The
sample is selected from images obtained with the Wide Field and
Planetary Camera 2 (WFPC2) on board the Hubble Space Telescope (HST)
taken from the HST-Archive.

The motivations of the project and the first results of this study
have been discussed discussed in the paper presenting the analysis of
the pilot sample (Treu et al.\ 1999; hereafter T99).  In this paper,
we describe in detail the selection of a larger sample of 35 objects,
and we present the surface photometry and the kinematic spectroscopic
measurements. An extended discussion of the state of the art in this
research area and of the objectives of our work is given in a
companion paper (Treu et al.\ 2001, submitted to MNRAS; hereafter
PIII), where the evolution of the Fundamental Plane and Kormendy
relation is analyzed.

This paper is organized as follows. The sample selection criteria and
the characteristics of the sample are discussed in Section
\ref{sec:sample}. In Section \ref{sec:photo}, we report on the
measurement of the photometric parameters. The measurement of
redshifts and central velocity dispersions is described in Section
\ref{sec:spec}. The caveats of the photometric and kinematic
measurements are extensively discussed. In particular, we discuss:
{\it i)} the accuracy that can be achieved in measuring the
combination of photometric observables that enters the Fundamental
Plane; {\it ii)} the effects of resolution on the measurement of
internal velocity dispersion, by means of Montecarlo simulations; {\it
iii)} the stability of the measured internal velocity dispersion with
respect to the rest frame spectral region used for the fit.  A summary
is given in Section~\ref{sec:sampdisc}.

The Hubble constant is assumed to be H$_0=50h_{50}$\kms
Mpc$^{-1}$. The matter density of the Universe and the cosmological
constant in dimesionless form are indicated as $\Omega$ an
$\Omega_{\Lambda}$ respectively.

\section{Sample selection}
\label{sec:sample}

\subsection{Selection criteria}

The targets used in this study have been chosen from a sample of
random early-type galaxies (E/S0) found in the WFPC2 parallel images
collected by the Medium Deep Survey (Griffiths et al.\ 1994). The
Medium Deep Survey consists of a database of 250 WFPC2 fields with at
least 1 image in each of the filters F606W and F814W, which broadly
correspond to Johnson-Cousins filters V and I; we refer to these
filters as \Vs\, and \Io.  The images have been reduced and analyzed
by the MDS group and the complete catalog is available to the
scientific community (Ratnatunga, Griffiths \& Ostrander 1999).

For the sample selection we used the following criteria:

\begin{enumerate}
\item {\bf Morphology clearly defined as early-type}. We selected
the galaxies classified by the MDS as pure bulges or disc plus bulge
objects with bulge-to-total luminosity (B/T) greater than 0.3 (the MDS
group considers \dv\, and exponential luminosity profiles, indicating
them as bulge and disc luminosity profile respectively; see the MDS
publications for details). All galaxies were inspected by eye to check
their morphology and some of them were rejected as mis-identifications
(stars or galaxies with clearly present spiral structure). We did not
exclude from the sample galaxies with nearby companions or disturbed
galaxies, in order not to bias against merging and interaction.
\item {\bf Sufficiently high luminosity}: (\Io~$<19.3$), which is the 
effective limit with EFOSC2 at the ESO-3.6m with the adopted grism \#9
in normally good seeing conditions ($0\farcs8-1''$).
\item {\bf High galactic latitude}: ($|b|>15^{\circ}$), 
to limit foreground star contamination and galactic extinction.
\item {\bf Low extinction}: ($\tx{E(B-V)}<0.2$), as measured by
Schlegel, Finkbeiner \& Davis (1998), to reduce the possibility of
photometric errors due to patchy reddening.
\item {\bf Appropriate colour}: ($0.95<$~\Vs-\Io~$<1.7$), which is the entire 
range spanned by E/S0 in the approximate redshift range $z=0.1-0.6$.
\item {\bf Clustering properties}. Parallel images are taken at random 
pointings within a few arcminutes of the primary target. Therefore,
when the primary target is a known cluster, the parallel image can
also contain part of the cluster. In order to avoid such
contamination, and to obtain an unbiased field sample, images centred
on known clusters were discarded from the analysis.
\end{enumerate}
\subsection{Discussion of the adopted selection criteria}

Apart from the natural requirements on the magnitude limit and the limits
on galactic latitude and foreground extinction, some selection
criteria deserve a thorough discussion. In fact, understanding the
selection criteria is a crucial step for a meaningful interpretation
of the data (e.g., see the discussion in Schade et al.\ 1999 on
morphology and colour selection criteria).

\begin{figure}
\mbox{\epsfysize=8cm \epsfbox{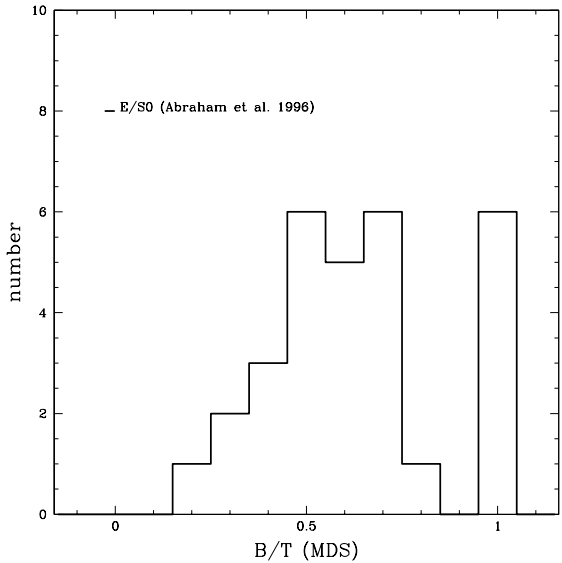}}
\caption{Comparison between the morphological classification of
Abraham et al.\ (1996) and our selection criteria.  The catalog of all
galaxies visually classified as E/S0 by Abraham et al.\ (1996) was
correlated with the MDS catalog by matching the absolute coordinates.
The distribution of the bulge-to-total luminosity ratio (B/T) measured
by the MDS for this catalog is plotted. A small value of the B/T cut
(we adopted B/T$>0.3$) is needed in order not to reject a significant
number of early-type galaxies.}
\label{fig:A2MDS}
\end{figure}

\begin{figure}
\mbox{\epsfysize=8cm \epsfbox{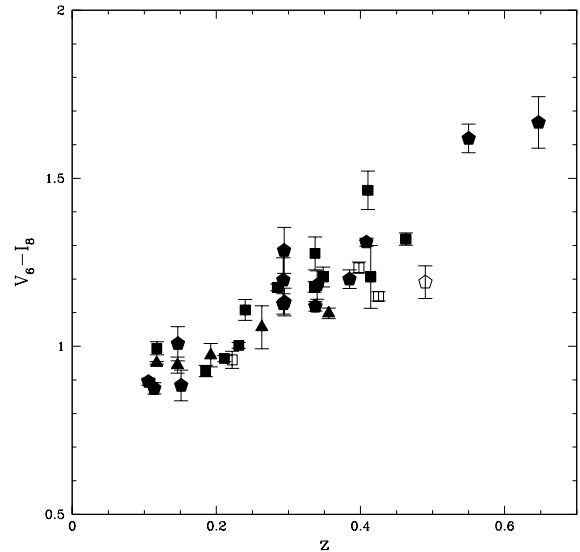}}
\caption{Observed \Vs-\Io\, colour (corrected for foreground
extinction) as a function of redshift for our sample. The galaxies are
plotted as pentagons (E), squares (E/S0) or triangles (S0) according
to their morphological classification. Morphological classification
based on HST morphology and surface photometry is discussed in
Section~\ref{ssec:morph}. Peculiar galaxies are identified by an open
symbol. In order of increasing redshift they are: {\bf F3} ($z=0.222$;
central dust lane, see Section~\ref{sec:photo}), {\bf E2, C2}
($z=0.398,0.425$ strong emission lines; see the spectrum of {\bf E2}
in Figure~\ref{fig:spec1}), and {\bf B3} ($z=0.490$; strong Balmer
absorption features, see Figure~\ref{fig:spec1}). Note that at any
given redshift morphologically classified objects span a significant
range in colour.}
\label{fig:colorz}
\end{figure}

\begin{description}

\item {\bf Morphological Selection}. We checked our morphological
selection criterion (B/T~$>0.3$) by studying the distribution of B/T
values for the galaxies morphologically classified as early-type by
Abraham et al.\ (1996). By matching absolute coordinates, we
cross-correlated the MDS and Abraham et al.\ catalogs down to
\Io~$=21$ recovering the B/T value assigned by the MDS to each galaxy
visually classified as an early-type by Abraham et al.\ (1996). The
distribution of B/T values found by the MDS group is plotted in
Figure~\ref{fig:A2MDS}. We conclude that our threshold in B/T does not
exclude {\it a priori} any significant part of the early-type galaxy
population.

\item {\bf Colour selection}. Our experimental setup was tuned to work in 
the redshift range $z=0.1-0.6$. Therefore, we chose very loose colour
cuts that at the same time could reject E/S0 outside our target
redshift range ($z<0.1$ or $z>0.6$) and include the widest range of
stellar population properties. The colour-redshift relation for our
sample is shown in Figure~\ref{fig:colorz}. It is noticed that at any
given redshift there are objects with early-type morphology that are
significantly bluer than the reddest ones. The effect of the adopted
colour selection criterion on the study of the evolution of stellar
populations is discussed in PIII.

\item {\bf Clustering Properties}. We did not attempt to identify and to 
exclude members of groups or poor clusters. In fact, four of the six
objects for which we have obtained FP parameters in T99 are likely to
be members of a group (or a poor cluster). Hence, the early-type
galaxies in our sample should be representative of a random pointing,
magnitude-selected, sample of objects.

\end{description}

\subsection{The sample}
\label{ssec:sample}

We took spectra of 25 galaxies satisfying the adopted colour and
magnitude selection criteria. For 22 of them the signal-to-noise ratio
achieved is sufficient to measure accurate velocity dispersions (see
Section~\ref{sec:spec}). By rotating the slit on the sky, we observed
spectra for 10 additional early-type galaxies as secondary
targets. For 6 secondary early-type galaxies the signal-to-noise ratio
achieved is sufficient to measure velocity dispersions. The redshift
has been measured for all the 35 galaxies for which we present
photometry.

In Section~\ref{sec:spec}, where we are mostly interested in
discussing the measurement of velocity dispersions, we will consider
all the galaxies including the secondary targets. In turn,
when analyzing the results of this study in terms of evolution of the
stellar populations of E/S0 (as in PIII), we will consider
only the galaxies satisfying the adopted selection criteria.

Table~\ref{tab:obj} lists the observed objects (the digit in the
adopted name indicates the observing run when the spectra were taken,
see Section~\ref{sec:spec}) with their coordinates and MDS
identification.  The objects are flagged to indicate whether they
satisfy all the selection criteria (sel=2), all criteria except that
they are fainter than the limit \Io~$=19.3$ (sel=1), or they are added
only as the best secondary target in the field (sel=0). For some of
the objects the MDS photometry was not available (galaxy {\bf D3} is
not considered by the MDS group because it is close to the edge of the
combined frame used, see Table~\ref{tab:phlog}; sel=nn), or was not
reliable (chip edge for galaxy {\bf Q3}, dust lane for galaxy {\bf
F3}; sel=-1). For easy reference Table~\ref{tab:obj} also lists for
each object the number of spectroscopic observations, the total
spectroscopic integration times, the S/N achieved, and the redshift of
the object.

\begin{table*}
\caption{Galaxy identification. For each galaxy we list the adopted
name (gal), the MDS field name and identification number within the
field (MDS ID), MDS coordinates (RA DEC), satisfaction of selection
criteria (sel, see Section~\ref{ssec:sample}), and whether $\sigma$
was measurable or not ($\sigma$). Galaxy {\bf D3} is not present in
the MDS catalog (see Section~\ref{ssec:sample}) and is marked with nn
in columns MDS~ID and sel. In addition, for each galaxy we list:
number of spectroscopic observations (n), total spectroscopic
exposures time (texp in seconds), and average S/N per pixel
(S/N). Finally, the redshift is listed in the last column, with four
digits if determined with the GHFF (see Section~4), with three if
determined by recognizing the most important absorption features.}
\label{tab:obj}
\begin{tabular}{c l c c c c c c c c }
gal	& MDS ID	& RA(J2000)	& DEC(J2000)	&  sel 	  & $\sigma$ & n & texp & S/N & z \\
\hline								
A1	& ut800\_1	& 14:42:14.72	& -17:11:43.7	& 2	  & y	   &  4   &  7200  &  65 & 0.1466      \\
B1	& ut800\_8	& 14:42:12.87	& -17:11:57.8	& 0	  & y	   &  4   &  7200  &  40 &  0.146      \\
D1	& ur610\_6	& 13:25:40.81	& -29:54:13.5	& 2	  & y	   &  4   & 12000  &  22 & 0.3848      \\
E1	& u5405\_62	& 17:57:34.29	& +04:51:32.8	& 2	  & y	   &  5   & 13200  &  17 & 0.2940      \\
F1	& u5405\_91	& 17:57:34.58	& +04:51:42.4	& 2	  & n	   &  5   & 13200  &  12 & 0.2935      \\
G1	& u5405\_18	& 17:57:33.54	& +04:51:42.1	& 2	  & y	   &  3   & 10800  &  35 & 0.2946      \\
I1	& u5405\_47	& 17:57:33.47	& +04:52:07.3	& 2	  & y	   &  3   & 10800  &  22 & 0.2929      \\
O1	& ust00\_15	& 10:05:47.23	& -07:42:05.9	& 0	  & n	   & 10   & 36000  &  7  & 0.647       \\
P1	& ust00\_16	& 10:05:47.89	& -07:41:53.1	& 1	  & n	   & 10   & 36000  &  6  & 0.550       \\
B2	& upu00\_13	& 09:59:18.54	& -22:54:46.5	& 1	  & n	   &  2   &  7200  &  6  & 0.463       \\
C2	& upz00\_10	& 13:22:09.46	& -36:40:10.9	& 2	  & n	   &  2   &  7200  &  9  & 0.425       \\
D2	& urw10\_3	& 11:21:24.02	& -24:55:18.6	& 2	  & y	   &  2   &  7200  &  33 & 0.1923      \\
E2	& uu301\_1	& 15:56:33.17	& +11:08:25.9	& 2	  & y	   &  4   & 13500  &  24 & 0.3977      \\
F2	& uu301\_5	& 15:56:32.86	& +11:07:57.6	& 2	  & y	   &  4   & 13500  &  18 & 0.3364      \\
A3	& ux100\_2	& 14:45:09.52	& +10:01:37.5	& 2	  & y	   &  3   &  7200  &  19 & 0.2311      \\
B3	& ux100\_5	& 14:45:09.76	& +10:02:09.7	& 2	  & n	   &  3   &  7200  &  12 &  0.490      \\
C3	& ud200\_2	& 21:58:40.34	& -30:22:32.4	& 2	  & y	   &  2   &  3600  &  27 & 0.1057      \\
D3	& uha01\_nn 	& 21:32:32.79	& +00:15:04.8	& nn 	  & y	   &  2   &  3600  &  16 & 0.1511      \\
E3	& uha01\_12	& 21:32:34.28	& +00:14:13.8	& 2	  & y	   &  4   &  10800 &  17 & 0.2631      \\
F3	& uha01\_10	& 21:32:33.89	& +00:14:36.1	& -1	  & n	   &  2   &  7200  &  10 & 0.2220      \\
G3	& ubz09\_2	& 00:49:35.21	& -52:03:58.7	& 2	  & y	   &  7   &  23400 &  21 & 0.4081      \\
I3	& uci10\_2	& 01:24:39.56	& +03:52:21.4	& 2	  & y	   &  7   &  25200 &  22 & 0.4103      \\
L3	& uci10\_8	& 01:24:38.07	& +03:52:06.9	& 1	  & y	   &  7   &  25200 &  13 & 0.4141      \\
M3	& uv100\_8	& 15:06:20.65	& +01:44:02.6	& 2	  & y	   &  4   &  7200  &  16 & 0.3374      \\
Q3	& uv100\_1	& 15:06:21.45	& +01:45:16.7	& -1	  & y	   &  2   &  3600  &  41 & 0.1139      \\
R3	& ua-00\_2	& 01:02:24.52	& -27:10:39.9	& 2	  & y	   &  4   &  12400 &  24 & 0.3482      \\
S3	& ua-00\_3	& 01:02:22.78	& -27:10:58.7	& 2	  & y	   &  5   &  16000 &  18 & 0.3559      \\
T3	& ucs01\_2	& 02:56:20.69	& -33:21:22.3	& 2	  & y	   &  2   &  1800  &  14 & 0.1174      \\
U3	& ucs01\_4	& 02:56:25.11	& -33:22:08.1	& 0	  & y	   &  4   &  7200  &  19 & 0.2401      \\
A4	& ucs01\_3	& 02:56:25.26	& -33:21:24.7	& 2	  & y	   &  2   &  3600  &  25 & 0.1171      \\
B4	& ua400\_3	& 00:24:56.52	& -27:16:10.2	& 2	  & y	   &  2   &  5356  &  19 & 0.2112      \\
D4	& ufr00\_2	& 04:08:38.28	& -24:23:53.3	& 2	  & y	   &  5   &  13567 &  32 & 0.2848      \\
E4	& ufr00\_3	& 04:08:38.94	& -24:25:44.4	& 0	  & y	   &  5   &  13567 &  27 & 0.1851      \\
H4	& uim03\_1	& 03:55:33.99	& +09:43:11.5	& 2	  & y	   &  7   &  25200 &  25 & 0.3399      \\
I4	& uim03\_7	& 03:55:35.09	& +09:43:32.6	& 2	  & y	   &  7   &  25200 &  19 & 0.3369      \\
\hline
\end{tabular}
\end{table*} 

\begin{figure}
\mbox{\epsfysize=8cm \epsfbox{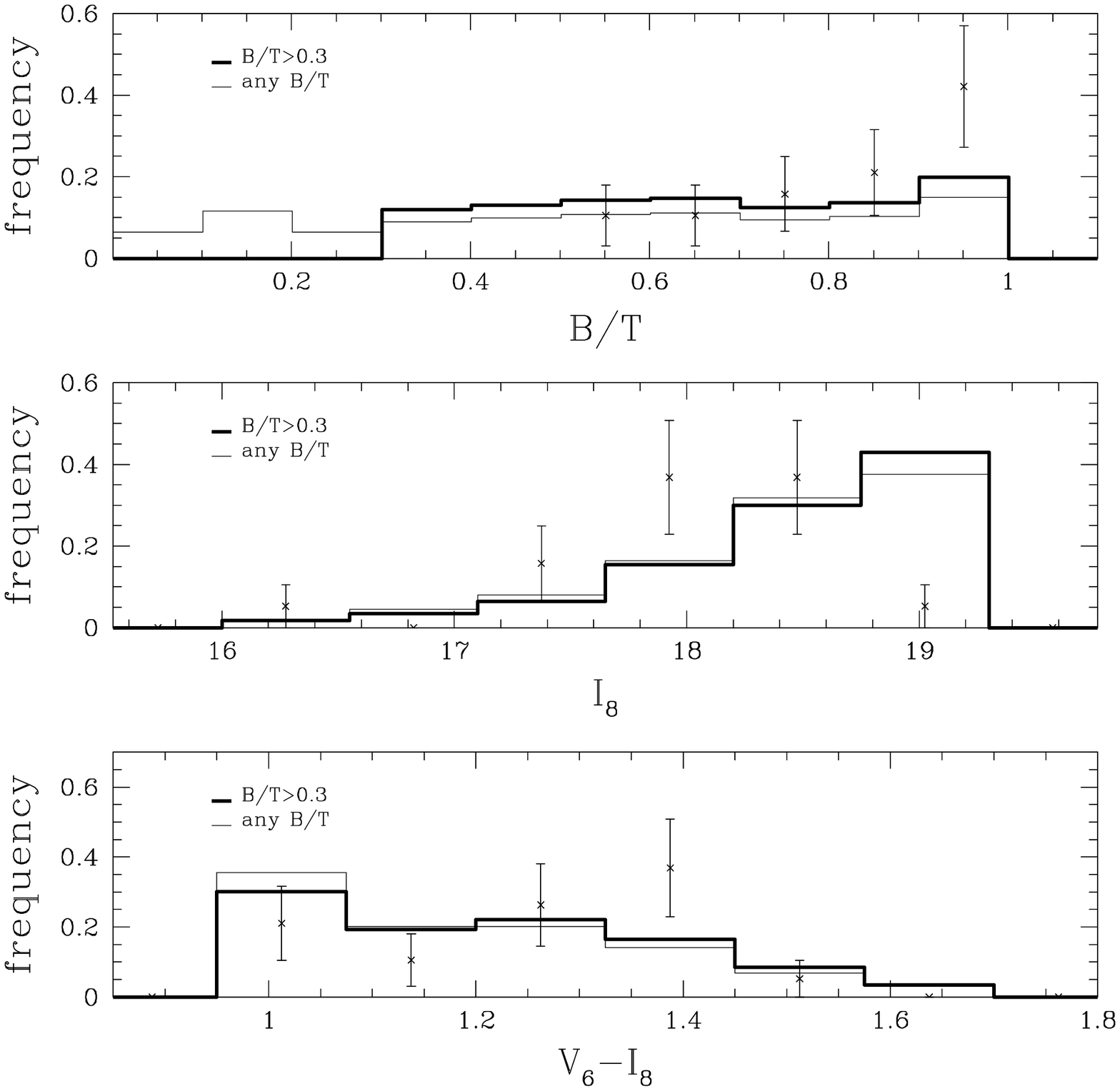}}
\caption{Comparison of the properties of the galaxies in our sample
with measured velocity dispersion to the ones of the galaxies in the
full MDS satisfying our colour and magnitude selection criteria
(\Io~$<19.3$, $0.95<$~\Vs-\Io~$<1.7$).  From top to bottom, we plot
the distribution of bulge-to-total luminosity ratio (B/T), magnitude
(\Io) and colour (\Vs-\Io) for our sample (points with error bars,
computed using Poisson statistics) and for the MDS catalog before and
after B/T cut (thin and thick histograms). We note that, as expected
since we avoided early-spirals, we have a marginally higher frequency
of large B/T values with respect to the entire catalog, and the small
B/T values are rejected. Furthermore, we have a smaller frequency of
objects in the faint magnitude bin, while no significant bias is
present in the colour distribution.}
\label{fig:bias3}
\end{figure}

Since we wish to use this sample to infer general properties of the
population of early-type galaxies, it is very important to understand
and quantify any bias that may be introduced in the selection process.
The distributions of apparent magnitude, colour, and B/T of the
galaxies with measured velocity dispersion and satisfying the relevant
selection criteria are shown in Figure~\ref{fig:bias3}, compared to
those of the galaxies with \Io~$<19.3$ and $0.95<$~\Vs-\Io~$<1.7$ in
the MDS. As expected from our morphological classification we have a
higher frequency of objects with large B/T. No major bias is found to
be present in the colour distribution.

\begin{figure}
\mbox{\epsfysize=8cm \epsfbox{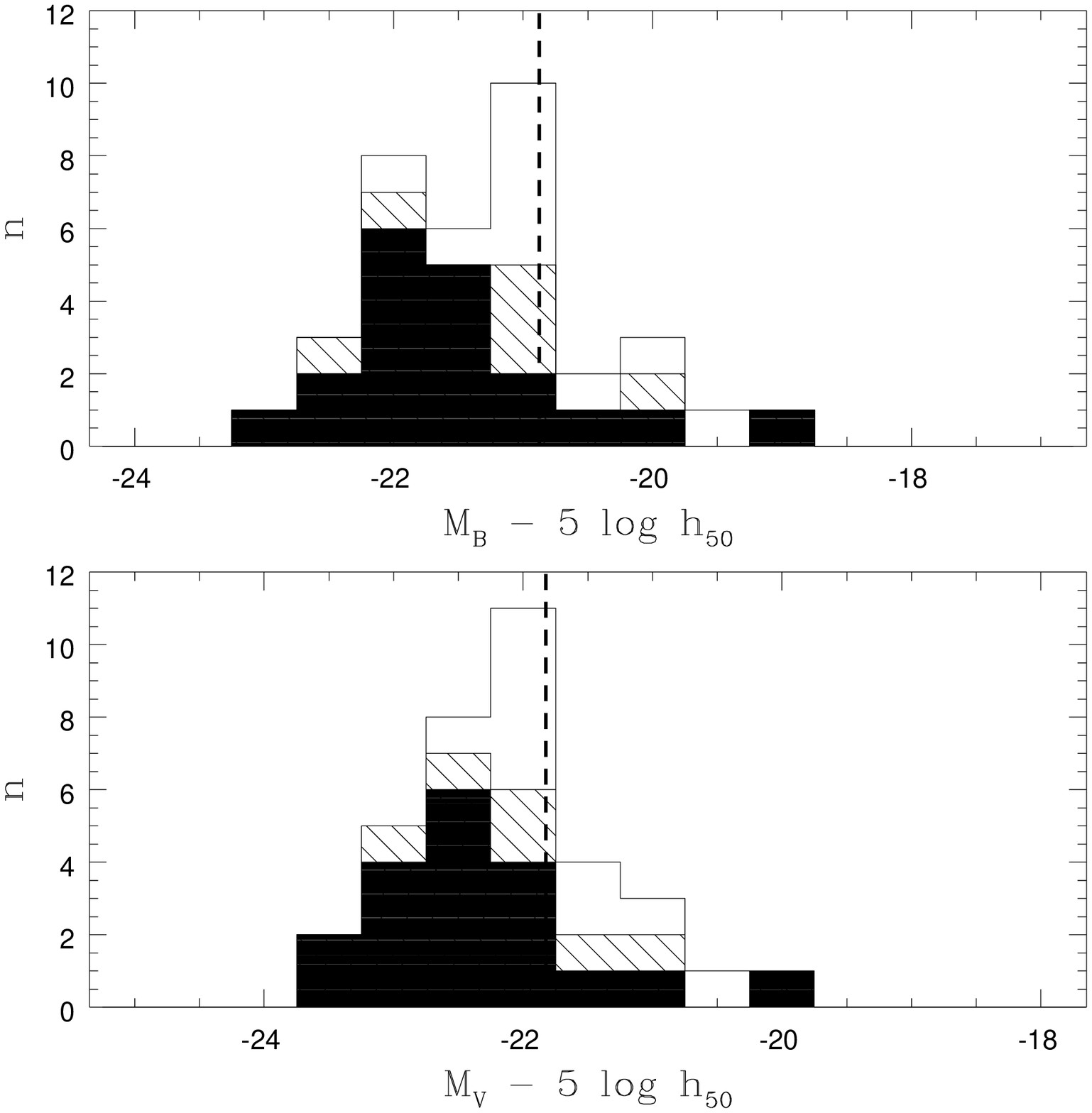}} 
\caption{Distribution of absolute magnitude for the galaxies
satisfying the colour and magnitude selection criteria (empty
histogram). The subset with measurable $\sigma$ ($\tx{S/N}>12$) is
plotted as hatched histogram.  The subset with $\sigma$ above
resolution limits (see Section~\ref{ssec:kineres}) is plotted as
filled histogram. The distributions for B and V rest frame
luminosities are shown in the upper and lower panel. The thick
vertical dashed line represents the characteristic magnitude
${\mathcal M}_{*}$ as measured by Marzke et al.\ (1998) in the local
Universe (in the V band ${\mathcal M}_{*}$ is obtained by assuming
$\tx{B-V}=0.95$, the colour of a single-burst stellar population, age
of 12 Gyr, solar metallicity, and Salpeter IMF (Salpeter 1955),
computed using Bruzual \& Charlot 1993, GISSEL96 version). The values
of the cosmological parameters $\Omega=1$ and $\Omega_{\Lambda}=0$ are
assumed.}
\label{fig:histoM}
\end{figure}

In addition it is important to notice that, because of the selection
effects (\Io~$<19.3$) and resolution limits (see
Section~\ref{ssec:kineres}), our sample is limited to relatively
massive luminous early-type galaxies. The distribution of absolute
magnitudes is plotted in Figure~\ref{fig:histoM}, together with the
location of the characteristic magnitude ${\mathcal M}_{*}$ of the
Schechter luminosity function (Schechter 1976).  The bulk of our
sample of galaxies is comprised between ${\mathcal M}_{*}$ and
${\mathcal M}_{*}-1$.

\section{Photometry}
\label{sec:photo}

The images are taken from the Medium Deep Survey of the Hubble Space
Telescope (Griffiths et al.\ 1994). For each MDS field there are
images from WFPC2 through filters \Vs\, and \Io. Table~\ref{tab:phlog}
lists total exposure times and number of exposures.
\begin{table}
\caption{Photometric data: number of exposures (in each of the two filters, 
n$_6$ and n$_8$) and total exposure times (in each of the two filters,
texp$_6$ and texp$_8$, in seconds). The fields uha00 and uha01 and
uim01 and uim03 overlap partially. Some of the fields have been imaged
with large dithering and therefore the exposure time is not uniform
across the field. When two galaxies in the same field have been imaged
with different exposure times we list the field twice, specifying the
two exposure times. In column note we specify which galaxy has been
imaged with the listed exposure times. Column E(B-V) lists the
galactic extinction as given by Schlegel et al.\ (1998).}
\label{tab:phlog}
\begin{tabular}{lrrrrlr}
field 	&  n$_6$ & texp$_6$	&  n$_8$	&	texp$_8$ & note & E(B-V)\\
\hline								  
ua400	& 4	&  8000		&	4	&	8000	 & 	& 0.014	\\
ua-00	& 2	&  2100		&	2	& 	4200	 & 	& 0.017	\\
ubz09	& 2 	&  2500		&	7	&	6400	 & 	& 0.019	\\
uci10	& 3	&  4800		& 	4	& 	10800	 & 	& 0.028	\\
ucs01	& 2	&  1500		&	2	&	4200	 & 	& 0.024	\\
ud200	& 1	&  1200		&	1	&	2100	 & 	& 0.021	\\
ufr00	& 3 	&  1900		&	3	&	2000	 & {\bf E4}	& 0.052	\\
ufr00	& 2 	&  1300		&	2	&	1400	 & {\bf D4}	& 0.052	\\
uha00-01& 5	&  5700		&	6	&	8400	 & {\bf E3,F3} & 0.044\\
uha00-01& 2	&  2600		&	3	&	4200	 & {\bf D3}	& 0.044	\\
uim01-03& 4	&  4700		&	3	&	4000	 & {\bf H4}	& 0.172	\\
uim01-03& 7	&  5900		&	7	&	5900	 & {\bf I4}	& 0.172	\\
upu00	& 2	&  2300		&	2	&	1700	 & 	& 0.055	\\
upz00	& 3	&  4100		&	3	&	2900	 & 	& 0.055	\\
urw10	& 2	&  2500		&	2	&	2000	 & 	& 0.043	\\
ur610	& 2 	&  1500 	&  	2 	&  	1600 	 & 	& 0.061	\\
ust00	& 10 	& 16500 	& 	11 	& 	23100 	 & 	& 0.048	\\
ut800	& 2 	&  1430 	&	3 	& 	4430 	 & 	& 0.099 \\
uu301	& 1	&  1000		&	3	&	2800	 & 	& 0.049	\\
uv100	& 2	&  4800		&	2	& 	3300	 & 	& 0.053	\\
ux100	& 4	&  1700		&	4	& 	1700	 & 	& 0.024	\\
u5405 	& 1 	&   800 	&	1 	&  	 800 	 & 	& 0.188	\\
\hline
\end{tabular} 
\end{table}

The data reduction procedure has been described in full detail in T99
(see also Treu 2001) and the description will not be repeated here.

For each galaxy we fitted isophotal profiles\footnote{Surface
brightness as a function of circularized radius, as defined in T99.},
and 2D models convolved with the Tiny Tim PSF (Krist 1994; see T99).
We define the photometric parameters, effective radius $r_{\tx{e}}$,
and effective surface brightness $SB_{\tx{e}}$, as the best-fitting
parameters of the \dv component. As in T99, we will denote the angular
effective radius (in arcseconds) by $r_{\tx{e}}$ and the linear
effective radius (in kiloparsecs) by $R_{\tx{e}}$.

The isophotal profiles were fitted with an \dv profile, an \dv\, plus
exponential profile, and an exponential profile. All galaxies in the
sample were best-fitted by the \dv\, or \dv plus exponential
luminosity profile.  Fifteen galaxies were best-fitted by a pure \dv\,
profile. Fifteen gave insignificantly smaller residual when an
exponential component was added (and negligible changes in the \dv\,
photometric parameters). For five galaxies the addition of the
exponential component improved significantly the fit (see
Section~\ref{ssec:morph}) and changed the parameters.  In the
following we adopt the values obtained by fitting a pure \dv\ profile,
in all cases except for the five galaxies for which the exponential is
needed (galaxies {\bf B1, D2, E3, S3, A4}), where the two components
are fitted and the parameters of the \dv\, component are adopted. The
2D models were pure \dv\, luminosity profiles, and the fit was
performed with two weighting schemes, least squares (2Dls) and least
$\chi^2$ (2Dl$\chi^2$).

The larger sample available for the study presented in this paper,
with respect to T99, allows us to perform a statistical analysis of
the photometric results. In the following we will give relative errors
in per cent when referred to effective radii, in magnitudes when
referred to magnitudes or surface brightnesses, and in absolute terms
when referred to logarithmic quantities.

The values obtained from luminosity profile fitting and 2Dl$\chi^2$
have an average difference of 15 per cent in $r_{\tx{e}}$ (in F606W; 14 per cent
in F814W) and r.m.s. scatter of 16
per cent (F606W and F814W), if we restrict the study to the objects
without a significant disky component. If the five objects with discs
are included the r.m.s increases to 53 per cent (F606W) and 37 per cent (F814W);
we emphasize that the 2D models have pure \dv\, luminosity profiles
and therefore an increased scatter is expected, because of the
different modelling. The least square fitting technique produces larger
differences and scatter in the results. Compared to the profile
fitting method, the average difference in $r_{\tx{e}}$ is 30 per cent
(F606W) and 28 per cent (F814W) with an r.m.s. scatter of 20 per cent (F606W) and
18 per cent (F814W). The r.m.s.  scatter increases to 32 per cent and 27 per cent if the
galaxies with a significant disc component are
considered. 

As noted by many authors, the effective radius and effective surface
brightness are correlated observables and the combination that enters
the FP,
\begin{equation}
FP_{ph}=\log r_{\tx{e}} - \beta SB_{\tx{e}},
\end{equation}
is particularly robust (see the careful discussion by Kelson et al.\
2000 of the results obtained by fitting a variety of models,
including the Sersic $r^{1/n}$ profile, with a variety of techniques,
including growth curves, luminosity profile fit and 2D fit). The
values of the quantity\footnote{The values listed here are computed
using $\beta=0.32$ (F606W) and $\beta=0.328$ (F814W).  However, since
the slope $\beta$ is very well determined (see Pahre, Djorgovski \& de
Carvalho 1998) and varies very little with wavelength, the estimate of
the errors changes negligibly within the observed range of values for
$\beta$.} $FP_{ph}$ obtained with the luminosity profile fit and
2Dl$\chi^2$ differ on average by -0.003 (F606W) and 0.001 (F814W) with
an r.m.s. scatter of 0.037 (F606W) and 0.034 (F814W), by including all
the galaxies.  In addition, $FP_{ph}$ is also remarkably stable for
2Dls; the average difference with respect to the results obtained by
fitting luminosity profiles is -0.029 (F606W) and -0.027 (F814W) with
an r.m.s. scatter of 0.031 (F606W) and 0.026 (F606W).

Because of the higher scatter and difference obtained with the 2Dls
method, we adopt as best estimate of the photometric observables the
average of the results obtained with the luminosity profile and the
2Dl$\chi^2$ method.  The best estimates of the photometric observables
are listed in Table~\ref{tab:photopar}, together with the uncertainty
taking into account errors related to sky subtraction, flatfielding
and fitting technique, computed as described in
Section~\ref{ssec:photoerr}. As a check, we compared the magnitudes we
derived here to those derived by the MDS group: the average difference
and r.m.s scatter are 0.02, 0.15 (\Vs ), 0.01, 0.12 (\Io ), and 0.01,
0.11 (\Vs$-$\Io ), after excluding galaxy {\bf F3} (see below). The
comparison takes into account the different photometric zero points
adopted by the MDS group (see the MDS web-site at URL
http://archive.stsci.edu/mds/) and the ones given on the HST User
Handbook that are adopted in the present paper.

\begin{table*}
\begin{center}
\caption{Observed photometric structural parameters. The uncertainty
on the combination that enters the FP, $FP_{ph}=\log
R_{\tx{e}}-\beta SB_{\tx{e}}$, is listed as $\delta FP$. The values
of $\delta FP$ listed here are calculated using $\beta=0.32$ (F606W)
and $\beta=0.328$ (F814W; see text).  The uncertainty on the colour is
listed as $\delta$68.}
\label{tab:photopar}
\begin{tabular}{lccccccccc}

gal	 & \Vs	 & $SB_{\tx{e6}}$  & r$_{\tx{e6}}$ & $\delta$FP$_{6}$	& \Io	 & $SB_{\tx{e8}}$	& r$_{\tx{e8}}$ & $\delta$FP$_8$ & $\delta$68 \\
\hline
A1 & $17.35 \pm	0.06 $& $20.39 \pm  0.06$ & $1.62 \pm  0.08$ & 0.011 & $16.25 \pm  0.10$ & $19.38 \pm  0.12$ & $1.70 \pm  0.17$ & 0.008 & 0.051\\
B1 & $19.00 \pm	0.02 $& $19.71 \pm  0.04$ & $0.56 \pm  0.01$ & 0.007 & $17.96 \pm  0.01$ & $18.66 \pm  0.04$ & $0.55 \pm  0.01$ & 0.007 & 0.024\\
D1 & $19.53 \pm	0.11 $& $20.98 \pm  0.20$ & $0.78 \pm  0.11$ & 0.004 & $18.28 \pm  0.08$ & $19.44 \pm  0.16$ & $0.69 \pm  0.08$ & 0.005 & 0.027\\
E1 & $20.03 \pm	0.06 $& $21.79 \pm  0.14$ & $0.90 \pm  0.08$ & 0.008 & $18.56 \pm  0.10$ & $20.41 \pm  0.19$ & $0.94 \pm  0.12$ & 0.004 & 0.068\\
F1 & $20.37 \pm	0.03 $& $22.58 \pm  0.03$ & $1.10 \pm  0.00$ & 0.010 & $19.07 \pm  0.02$ & $21.11 \pm  0.02$ & $1.02 \pm  0.00$ & 0.007 & 0.030\\
G1 & $18.86 \pm	0.09 $& $21.83 \pm  0.19$ & $1.57 \pm  0.19$ & 0.007 & $17.55 \pm  0.07$ & $20.36 \pm  0.15$ & $1.46 \pm  0.15$ & 0.007 & 0.041\\
I1 & $19.99 \pm	0.13 $& $20.98 \pm  0.25$ & $0.64 \pm  0.11$ & 0.007 & $18.61 \pm  0.10$ & $19.60 \pm  0.23$ & $0.63 \pm  0.09$ & 0.010 & 0.067\\
O1 & $21.78 \pm	0.14 $& $22.97 \pm  0.31$ & $0.70 \pm  0.13$ & 0.014 & $20.07 \pm  0.07$ & $20.85 \pm  0.29$ & $0.58 \pm  0.09$ & 0.022 & 0.077\\
P1 & $21.67 \pm	0.08 $& $22.21 \pm  0.27$ & $0.52 \pm  0.08$ & 0.020 & $20.01 \pm  0.04$ & $20.48 \pm  0.19$ & $0.50 \pm  0.05$ & 0.019 & 0.042\\
B2 & $21.36 \pm	0.08 $& $21.20 \pm  0.29$ & $0.38 \pm  0.06$ & 0.024 & $19.99 \pm  0.07$ & $20.22 \pm  0.22$ & $0.45 \pm  0.06$ & 0.014 & 0.018\\
C2 & $19.48 \pm	0.08 $& $21.57 \pm  0.16$ & $1.05 \pm  0.12$ & 0.002 & $18.28 \pm  0.08$ & $20.25 \pm  0.12$ & $0.99 \pm  0.09$ & 0.000 & 0.014\\
D2 & $18.70 \pm	0.05 $& $18.72 \pm  0.96$ & $0.39 \pm  0.23$ & 0.012 & $17.68 \pm  0.04$ & $17.94 \pm  0.74$ & $0.42 \pm  0.20$ & 0.015 & 0.035\\
E2 & $18.94 \pm	0.04 $& $21.99 \pm  0.08$ & $1.63 \pm  0.09$ & 0.004 & $17.66 \pm  0.03$ & $20.47 \pm  0.10$ & $1.46 \pm  0.09$ & 0.006 & 0.016\\
F2 & $19.61 \pm	0.06 $& $21.60 \pm  0.13$ & $1.00 \pm  0.09$ & 0.004 & $18.39 \pm  0.05$ & $20.19 \pm  0.10$ & $0.92 \pm  0.07$ & 0.003 & 0.016\\
A3 & $18.75 \pm	0.08 $& $20.82 \pm  0.14$ & $1.04 \pm  0.10$ & 0.001 & $17.72 \pm  0.08$ & $19.41 \pm  0.14$ & $0.87 \pm  0.09$ & 0.005 & 0.009\\
B3 & $20.36 \pm	0.13 $& $19.18 \pm  0.18$ & $0.23 \pm  0.01$ & 0.048 & $19.15 \pm  0.08$ & $17.96 \pm  0.31$ & $0.23 \pm  0.02$ & 0.057 & 0.049\\
C3 & $18.86 \pm	0.11 $& $19.20 \pm  0.14$ & $0.47 \pm  0.05$ & 0.008 & $17.94 \pm  0.11$ & $18.22 \pm  0.10$ & $0.45 \pm  0.04$ & 0.010 & 0.011\\
D3 & $19.44 \pm	0.15 $& $19.54 \pm  0.44$ & $0.43 \pm  0.10$ & 0.027 & $18.51 \pm  0.13$ & $18.45 \pm  0.41$ & $0.40 \pm  0.10$ & 0.029 & 0.046\\
E3 & $19.59 \pm	0.10 $& $22.16 \pm  0.12$ & $1.30 \pm  0.13$ & 0.006 & $18.49 \pm  0.10$ & $21.03 \pm  0.16$ & $1.29 \pm  0.15$ & 0.003 & 0.064\\
F3 & $18.95 \pm	0.11 $& $21.86 \pm  0.28$ & $1.54 \pm  0.28$ & 0.018 & $17.95 \pm  0.09$ & $20.81 \pm  0.33$ & $1.51 \pm  0.29$ & 0.027 & 0.025\\
G3 & $19.85 \pm	0.14 $& $21.36 \pm  0.36$ & $0.82 \pm  0.19$ & 0.016 & $18.52 \pm  0.13$ & $19.87 \pm  0.40$ & $0.77 \pm  0.18$ & 0.025 & 0.012\\
I3 & $20.08 \pm	0.17 $& $20.98 \pm  0.28$ & $0.62 \pm  0.12$ & 0.003 & $18.59 \pm  0.13$ & $19.63 \pm  0.22$ & $0.65 \pm  0.11$ & 0.003 & 0.057\\
L3 & $20.85 \pm	0.15 $& $21.87 \pm  0.26$ & $0.65 \pm  0.11$ & 0.008 & $19.61 \pm  0.14$ & $20.31 \pm  0.35$ & $0.56 \pm  0.12$ & 0.018 & 0.093\\
M3 & $19.39 \pm	0.14 $& $21.13 \pm  0.18$ & $0.90 \pm  0.13$ & 0.006 & $18.22 \pm  0.13$ & $19.73 \pm  0.13$ & $0.81 \pm  0.10$ & 0.010 & 0.014\\
Q3 & $17.38 \pm	0.06 $& $20.22 \pm  0.09$ & $1.48 \pm  0.11$ & 0.002 & $16.45 \pm  0.06$ & $19.16 \pm  0.06$ & $1.39 \pm  0.07$ & 0.002 & 0.017\\
R3 & $19.09 \pm	0.05 $& $21.48 \pm  0.12$ & $1.20 \pm  0.09$ & 0.005 & $17.87 \pm  0.07$ & $20.14 \pm  0.21$ & $1.15 \pm  0.15$ & 0.011 & 0.029\\
S3 & $19.70 \pm	0.03 $& $22.12 \pm  0.29$ & $1.17 \pm  0.11$ & 0.051 & $18.59 \pm  0.02$ & $20.09 \pm  0.07$ & $0.77 \pm  0.05$ & 0.007 & 0.016\\
T3 & $18.00 \pm	0.05 $& $20.49 \pm  0.20$ & $1.25 \pm  0.09$ & 0.032 & $16.99 \pm  0.06$ & $19.71 \pm  0.07$ & $1.40 \pm  0.04$ & 0.019 & 0.019\\
U3 & $18.96 \pm	0.04 $& $20.61 \pm  0.09$ & $0.85 \pm  0.02$ & 0.020 & $17.83 \pm  0.07$ & $19.45 \pm  0.11$ & $0.84 \pm  0.01$ & 0.029 & 0.031\\
A4 & $18.24 \pm	0.03 $& $20.33 \pm  0.02$ & $0.95 \pm  0.12$ & 0.047 & $17.26 \pm  0.03$ & $19.20 \pm  0.13$ & $0.89 \pm  0.16$ & 0.035 & 0.004\\
B4 & $19.73 \pm	0.05 $& $20.52 \pm  0.05$ & $0.58 \pm  0.03$ & 0.007 & $18.75 \pm  0.06$ & $19.97 \pm  0.03$ & $0.70 \pm  0.03$ & 0.007 & 0.009\\
D4 & $18.99 \pm	0.05 $& $21.25 \pm  0.05$ & $1.13 \pm  0.05$ & 0.006 & $17.76 \pm  0.05$ & $19.96 \pm  0.06$ & $1.10 \pm  0.05$ & 0.005 & 0.010\\
E4 & $19.12 \pm	0.09 $& $20.19 \pm  0.05$ & $0.65 \pm  0.04$ & 0.012 & $18.14 \pm  0.08$ & $19.05 \pm  0.02$ & $0.61 \pm  0.02$ & 0.018 & 0.017\\
H4 & $19.34 \pm	0.07 $& $22.42 \pm  0.08$ & $1.65 \pm  0.12$ & 0.005 & $18.00 \pm  0.05$ & $20.95 \pm  0.06$ & $1.55 \pm  0.08$ & 0.002 & 0.042\\
I4 & $20.50 \pm	0.18 $& $21.48 \pm  0.38$ & $0.65 \pm  0.16$ & 0.015 & $19.06 \pm  0.15$ & $20.10 \pm  0.22$ & $0.65 \pm  0.11$ & 0.005 & 0.049\\
\hline
\end{tabular}
\end{center}
\end{table*} 

Galaxy {\bf F3} required a special treatment, which is worth an
extended discussion. The central part of the galaxy hosts a dust lane,
which alters significantly the luminosity profile.  In order to
recover the underlying surface luminosity we performed the following
correction. Assuming that the dust lane effects can be neglected at
large radii and that colour gradients are negligible for the
correction, we computed the observer frame extinction map,
\begin{equation}
E68(x,y)=\mu_6(x,y)-\mu_8(x,y)-\mu_{68}(\infty)
\end{equation}
where $\mu_6(x,y)$ and $\mu_8(x,y)$ are the surface brightness at any
given pixel through filters F606W and F814W and $\mu_{68}(\infty)$ is
the colour at large radii. Based on the extinction law given by
Cardelli, Clayton \& Mathis (1989) and on the redshift $z=0.222$, we
calculated the relations between the extinction map and the absorption
in the individual filters ($A_6$ and $A_8$):
\begin{equation}
A_6(x,y)=3.401 E68(x,y)
\end{equation}
\begin{equation}
A_8(x,y)=2.401 E68(x,y).
\end{equation}
The results of the correction are good as it can be judged from the
quality of the luminosity profile shown in Figure~\ref{fig:prof2} and
of the images shown in Appendix~\ref{app:2D}. However, because of this
correction, the galaxy is flagged by an open symbol in
Figure~\ref{fig:colorz}.

\begin{figure*}
\mbox{\epsfysize=18cm \epsfbox{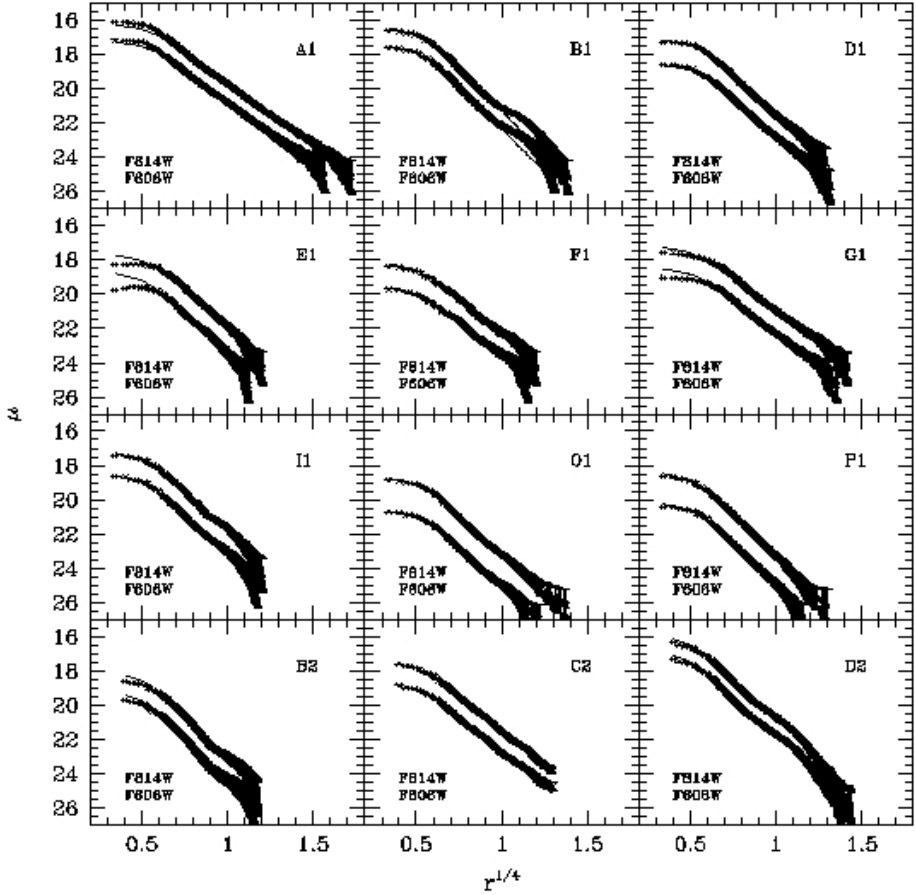}}
\caption{Luminosity profiles, I. The luminosity profiles obtained by
isophotal fitting are plotted together with the best-fitting \dv model
convolved with the PSF. Surface brightness ($\mu$) is plotted in mag
arcsec$^{-2}$, radius (r) in arcsec. The upper curves represent
luminosity profiles through filter F814W, the lower through filter
F606W.  The profiles of galaxies {\bf A1, B1, D1, I1, G1, E1, F1, O1,
P1} are taken from T99.  Galaxy {\bf B1} shows a spiral pattern in the
residuals of the two-dimensional fit (see T99). Galaxy {\bf D2} is
fitted by an \dv plus exponential model.}
\label{fig:prof1}
\end{figure*}

\begin{figure*}
\mbox{\epsfysize=18cm \epsfbox{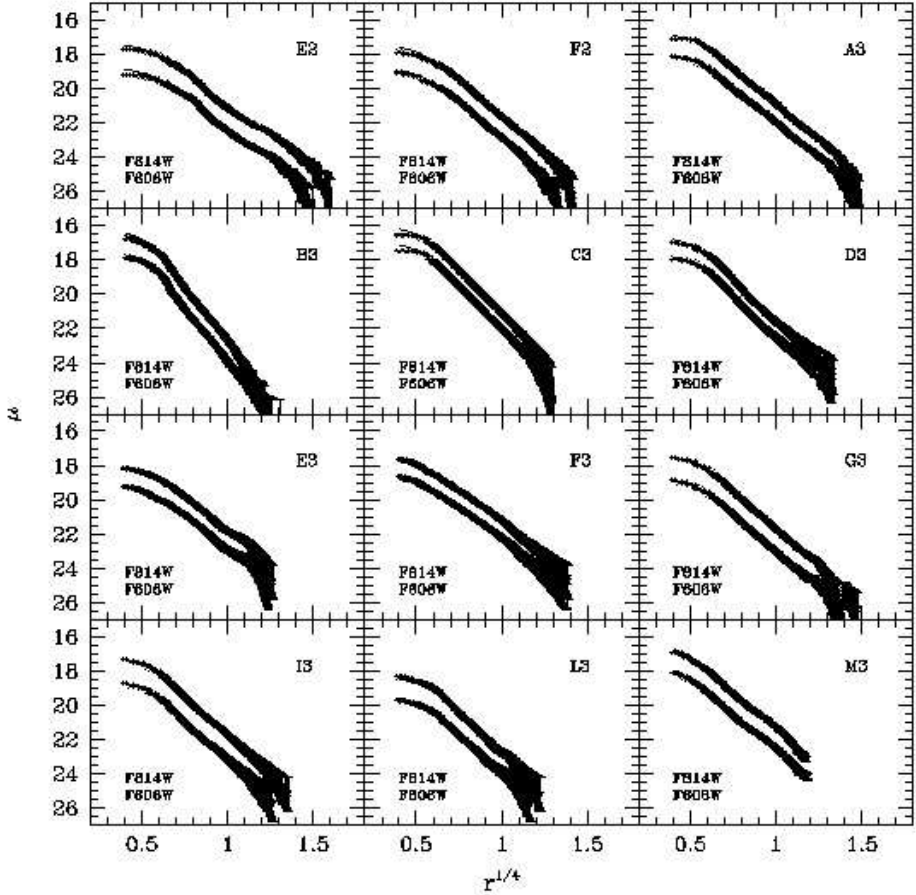}}
\caption{Luminosity profiles, II. As in Figure~\ref{fig:prof1}.  Galaxy
{\bf F3} hosts a dust lane in the central region; the profile shown
here is corrected for extinction as described in
Section~\ref{sec:photo}. Galaxy {\bf E3} has a clear spiral pattern,
see Figure~\ref{fig:2D2}.}
\label{fig:prof2}
\end{figure*}

\begin{figure*}
\mbox{\epsfysize=18cm \epsfbox{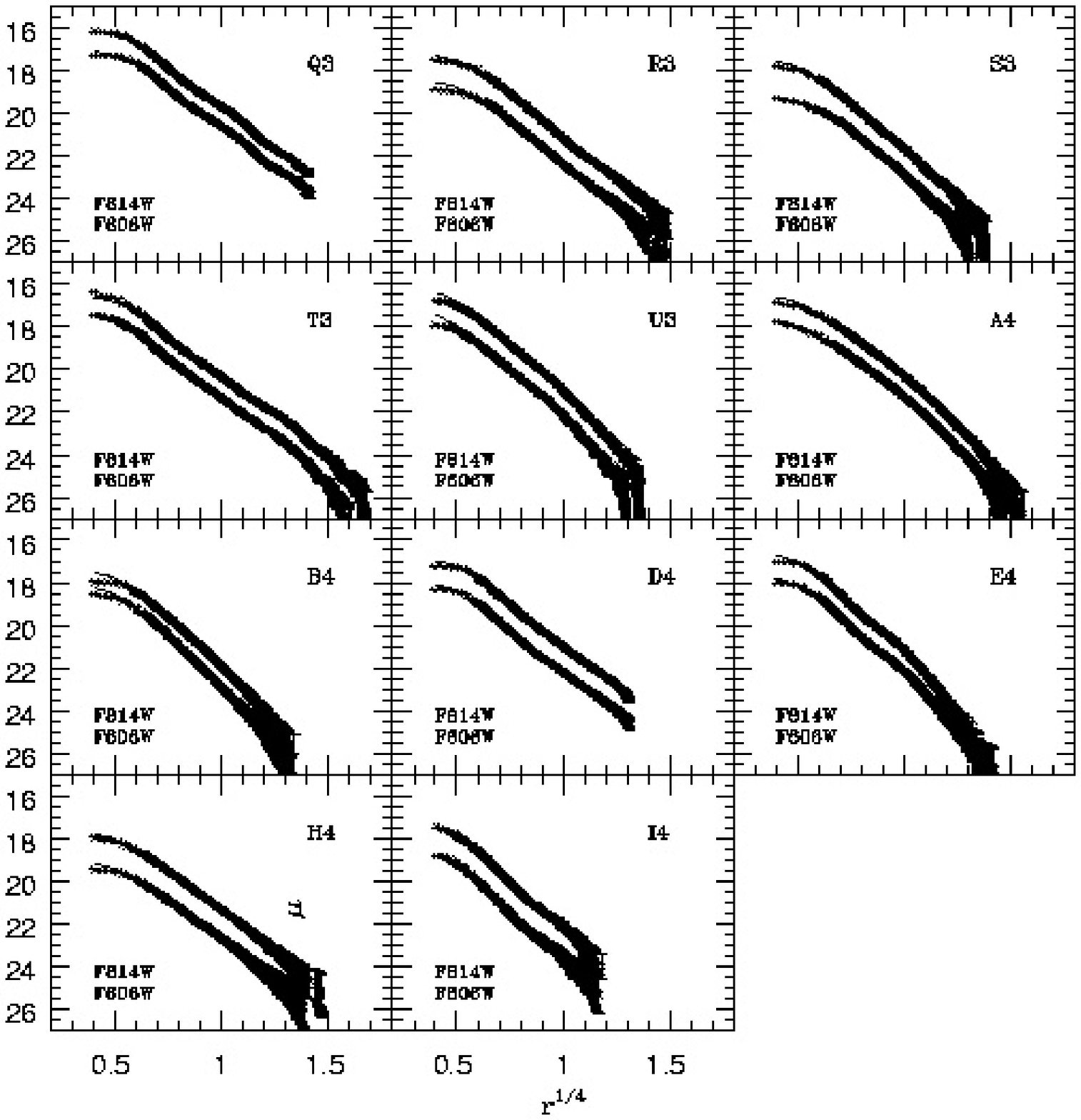}}
\caption{Luminosity profiles, III. As in Figure~\ref{fig:prof1}.
Galaxies {\bf S3} and {\bf A4} are fitted by an \dv plus exponential
model. }
\label{fig:prof3}
\end{figure*}

\subsection{Galactic extinction}

We used E(B-V) values given by Schlegel et al.\ (1998) and the
relations $A_6=2.889\, \tx{E(B-V)}$ and $A_8=1.948\, \tx{E(B-V)}$
calculated by Schlegel et al.\ (1998). The galactic extinction for
each field is listed in Table~\ref{tab:phlog}.  We compared the
extinction given by Schlegel et al.\ (1998) with the values given by
Burstein \& Heiles (1982). The average difference in E(B-V) is 0.017,
consistent with the different zero point used by the authors (see
Schlegel et al.\ 1998), and the scatter is 0.012.

\subsection{Rest frame photometric quantities}

\label{ssec:restphoto}

As in T99, the rest frame photometric quantities are computed from the
observed ones using the K-color correction ($\Delta$m$_{V8}$,
$\Delta$m$_{B6}$; listed in Table~\ref{tab:Kcolor}) defined as:
\begin{eqnarray}
\Delta m_{V8}\equiv&-2.5\log\Big[\int F_{\lambda}(\lambda)S_{V}(\lambda)d\lambda \Big]+\tx{zp}_{V}\nonumber\\&
+2.5\log\Big[\int F_{\lambda} (\lambda)S_{\tx{8}}(\lambda (1+z))d\lambda \Big]-\tx{zp}_{8},
\end{eqnarray} 
\begin{eqnarray}
\Delta m_{B6}\equiv & -2.5\log\Big[\int F_{\lambda}(\lambda)S_{B}(\lambda)d\lambda
\Big]+\tx{zp}_{B}\nonumber\\&+2.5\log\Big[\int F_{\lambda} (\lambda)S_{\tx{6}}(\lambda (1+z))d\lambda
\Big]-\tx{zp}_{6},
\end{eqnarray} 
\noindent
where zp and $S(\lambda)$ are the zero points and transmissions of the
band passes in the Landolt system, while $F_{\lambda}(\lambda)$ is the
flux per unit wavelength of the model spectrum. The model spectrum
used is the synthetic spectrum (from Bruzual \& Charlot 1993; GISSEL96
version) of a single-burst stellar population that best reproduces the
observed colour (see T99 and Section~\ref{ssec:photoerr}).

\begin{table}
\caption{K-color correction, computed as described in Section~\ref{ssec:restphoto}. Errors ($\delta\Delta m_{B6}$ and
$\delta\Delta m_{V8}$) are computed as described in
Section~\ref{ssec:photoerr}.}
\label{tab:Kcolor}
\begin{tabular}{lcccc}
gal & $\Delta m_{B6}$ & $\delta \Delta m_{B6}$ &
$\Delta m_{V8}$ & $\delta\Delta m_{V8}$ \\ 
\hline
A1 & 1.03 & 0.05 & 1.07 & 0.04 \\
B1 & 0.97 & 0.02 & 1.02 & 0.02 \\
D1 & 0.41 & 0.01 & 0.85 & 0.01 \\
E1 & 0.66 & 0.02 & 0.99 & 0.02 \\
F1 & 0.62 & 0.01 & 0.93 & 0.01 \\
G1 & 0.62 & 0.01 & 0.93 & 0.01 \\
I1 & 0.64 & 0.02 & 0.96 & 0.02 \\
O1 & -0.29 & 0.04 & 0.65 & 0.03 \\ 
P1 & -0.06 & 0.02 & 0.73 & 0.01 \\ 
B2 & 0.23 & 0.01 & 0.80 & 0.01 \\
C2 & 0.33 & 0.01 & 0.82 & 0.01 \\
D2 & 0.85 & 0.03 & 0.98 & 0.02 \\
E2 & 0.38 & 0.01 & 0.85 & 0.01 \\
F2 & 0.53 & 0.01 & 0.91 & 0.01 \\
A3 & 0.75 & 0.01 & 0.95 & 0.01 \\
B3 & 0.19 & 0.01 & 0.80 & 0.01 \\
C3 & 1.04 & 0.01 & 1.04 & 0.01 \\
D3 & 0.89 & 0.05 & 0.97 & 0.04 \\
E3 & 0.68 & 0.03 & 0.94 & 0.03 \\
F3 & 0.75 & 0.02 & 0.94 & 0.01 \\
G3 & 0.36 & 0.01 & 0.84 & 0.01 \\
I3 & 0.33 & 0.01 & 0.85 & 0.01 \\
L3 & 0.35 & 0.01 & 0.83 & 0.01 \\
M3 & 0.52 & 0.01 & 0.89 & 0.01 \\
Q3 & 0.99 & 0.02 & 1.01 & 0.01 \\
R3 & 0.50 & 0.01 & 0.90 & 0.01 \\
S3 & 0.47 & 0.01 & 0.87 & 0.01 \\
T3 & 1.12 & 0.02 & 1.12 & 0.02 \\
U3 & 0.78 & 0.02 & 0.99 & 0.02 \\
A4 & 1.07 & 0.01 & 1.08 & 0.01 \\
B4 & 0.78 & 0.01 & 0.95 & 0.01 \\
D4 & 0.66 & 0.01 & 0.96 & 0.01 \\
E4 & 0.83 & 0.01 & 0.96 & 0.01 \\
H4 & 0.52 & 0.01 & 0.90 & 0.01 \\
I4 & 0.54 & 0.01 & 0.93 & 0.01 \\
\hline
\end{tabular}
\end{table}

The radii $r_{\tx{eB}}$ and $r_{\tx{eV}}$ are calculated for the
central rest wavelength of each filter B and V by linear
interpolation/extrapolation in wavelength between the measured radii
in the WFPC2 filters F606W and F814W at their observed central
wavelengths.  We take the central wavelengths of B, V, F606W, and
F814W to be 4400, 5500, 5935, and 7921~\AA, respectively, resulting in
the following formulae for the angular sizes:
\begin{eqnarray}
 r_{\tx{e}{\sc B}}=&\big\{r_{\tx{e}{\sc F606W}}[7921-(1+z)4400]\nonumber\\
 &+r_{\tx{e}{\sc F814W}}[(1+z)4400-5935]\big\}/1986 \\
 r_{\tx{e}{\sc V}}=&\big\{r_{\tx{e}{\sc F606W}}[7921-(1+z)5500]\nonumber\\ 
 &+r_{\tx{e}{\sc F814W}}[(1+z)5500-5935]\big\}/1986 \, .
\end{eqnarray}
The rest frame surface brightness is computed as:
\begin{equation}
\tx{SB}_{\tx{e}{V}}=\tx{SB}_{\tx{e}{8}}-10\log(1+z)+\Delta\tx{m}_{V8}-\tx{A}_8
\end{equation}
\begin{equation}
\tx{SB}_{\tx{e}{B}}=\tx{SB}_{\tx{e}{6}}-10\log(1+z)+\Delta\tx{m}_{B6}-\tx{A}_6
\end{equation}
Table~\ref{tab:photorest} lists for each galaxy the effective radius
in kpc, computed for a cosmological model with $\Omega=1$,
$\Omega_{\Lambda}=0$, $h_{50}=1$, and the effective surface
brightness.

\begin{table*}

\caption{Rest frame photometric structural parameters. Effective radii are in
$h_{50}^{-1}$kpc (for $\Omega=1$ and $\Omega_{\Lambda}=0$). The error
on the rest frame colour B-V is listed as $\delta$BV. The morphological
classification (Class; Section~\ref{ssec:morph}) is also
listed. Photometric structural parameters for S0 and S refer to the
\dv component. Errors are computed as described in
Section~\ref{ssec:photoerr}.}
\label{tab:photorest}
\begin{tabular}{lcccccccc}
gal	     & $SB_{\tx{eB}}$	& $R_{\tx{eB}}$ & $\delta$FP$_{B}$  & $SB_{\tx{eV}}$	& $R_{\tx{eV}}$ & $\delta$FP$_V$ & $\delta$BV & Class\\
\hline
A1 & $20.54 \pm	 0.08$ & $5.32 \pm  0.47$ & 0.050 & $19.67 \pm	0.13$ & $5.49 \pm  0.25$ & 0.035 & 0.083 & E    \\
B1 & $19.81 \pm	 0.06$ & $1.86 \pm  0.05$ & 0.047 & $18.90 \pm	0.07$ & $1.85 \pm  0.03$ & 0.032 & 0.039 & S   \\
D1 & $19.80 \pm	 0.20$ & $4.90 \pm  0.63$ & 0.029 & $18.76 \pm	0.16$ & $4.43 \pm  0.42$ & 0.020 & 0.028 & E    \\
E1 & $20.79 \pm	 0.15$ & $4.85 \pm  0.48$ & 0.087 & $19.92 \pm	0.21$ & $5.03 \pm  0.43$ & 0.059 & 0.074 & E    \\
F1 & $21.54 \pm	 0.07$ & $6.02 \pm  0.01$ & 0.087 & $20.56 \pm	0.09$ & $5.72 \pm  0.01$ & 0.059 & 0.033 & E    \\
G1 & $20.78 \pm	 0.20$ & $8.64 \pm  1.19$ & 0.087 & $19.80 \pm	0.18$ & $8.20 \pm  0.64$ & 0.059 & 0.045 & E    \\
I1 & $19.97 \pm	 0.26$ & $3.46 \pm  0.65$ & 0.087 & $19.07 \pm	0.24$ & $3.44 \pm  0.38$ & 0.060 & 0.073 & E    \\
O1 & $20.38 \pm	 0.31$ & $4.83 \pm  0.60$ & 0.030 & $19.24 \pm	0.29$ & $3.97 \pm  1.30$ & 0.028 & 0.092 & E    \\
P1 & $20.11 \pm	 0.27$ & $3.75 \pm  0.36$ & 0.030 & $19.22 \pm	0.19$ & $3.64 \pm  0.52$ & 0.024 & 0.048 & E    \\
B2 & $19.62 \pm	 0.29$ & $2.72 \pm  0.33$ & 0.035 & $19.26 \pm	0.22$ & $3.12 \pm  0.44$ & 0.022 & 0.019 & E/S0 \\
C2 & $20.21 \pm	 0.16$ & $6.91 \pm  0.65$ & 0.026 & $19.42 \pm	0.12$ & $6.59 \pm  0.57$ & 0.017 & 0.014 & E/S0 \\
D2 & $18.68 \pm	 0.96$ & $1.54 \pm  1.30$ & 0.025 & $18.08 \pm	0.74$ & $1.63 \pm  0.70$ & 0.021 & 0.050 & S0   \\
E2 & $20.78 \pm	 0.09$ & $10.34\pm  0.51$ & 0.023 & $19.77 \pm	0.10$ & $9.50 \pm  0.51$ & 0.017 & 0.017& E/S0 \\
F2 & $20.73 \pm	 0.14$ & $5.90 \pm  0.55$ & 0.023 & $19.74 \pm	0.11$ & $5.53 \pm  0.31$ & 0.016 & 0.016 & E/S0 \\
A3 & $20.59 \pm	 0.14$ & $5.05 \pm  0.61$ & 0.011 & $19.41 \pm	0.14$ & $4.51 \pm  0.33$ & 0.009 & 0.010 & E/S0 \\
B3 & $17.57 \pm	 0.18$ & $1.64 \pm  0.06$ & 0.049 & $16.98 \pm	0.31$ & $1.64 \pm  0.19$ & 0.057 & 0.050 & E    \\
C3 & $19.74 \pm	 0.14$ & $1.23 \pm  0.22$ & 0.013 & $18.78 \pm	0.10$ & $1.21 \pm  0.13$ & 0.012 & 0.021 & E    \\
D3 & $19.69 \pm	 0.44$ & $1.51 \pm  0.53$ & 0.038 & $18.73 \pm	0.41$ & $1.45 \pm  0.29$ & 0.034 & 0.078 & E    \\
E3 & $21.70 \pm	 0.12$ & $6.63 \pm  0.80$ & 0.023 & $20.87 \pm	0.16$ & $6.59 \pm  0.51$ & 0.016 & 0.074 & S   \\
F3 & $21.61 \pm	 0.29$ & $7.04 \pm  1.66$ & 0.027 & $20.79 \pm	0.33$ & $6.95 \pm  0.92$ & 0.030 & 0.033 & E/S0 \\
G3 & $20.17 \pm	 0.36$ & $5.27 \pm  1.06$ & 0.018 & $19.19 \pm	0.40$ & $5.00 \pm  1.10$ & 0.025 & 0.012 & E    \\
I3 & $19.74 \pm	 0.28$ & $4.04 \pm  0.70$ & 0.013 & $18.93 \pm	0.23$ & $4.24 \pm  0.64$ & 0.009 & 0.058 & E/S0 \\
L3 & $20.64 \pm	 0.26$ & $4.15 \pm  0.63$ & 0.016 & $19.58 \pm	0.35$ & $3.71 \pm  0.72$ & 0.020 & 0.094 & E/S0 \\
M3 & $20.24 \pm	 0.18$ & $5.31 \pm  0.79$ & 0.025 & $19.26 \pm	0.13$ & $4.90 \pm  0.47$ & 0.019 & 0.015 & E    \\
Q3 & $20.58 \pm	 0.10$ & $4.19 \pm  0.45$ & 0.025 & $19.60 \pm	0.07$ & $4.03 \pm  0.26$ & 0.017 & 0.028 & E    \\
R3 & $20.63 \pm	 0.12$ & $7.18 \pm  0.56$ & 0.009 & $19.71 \pm	0.21$ & $6.94 \pm  0.68$ & 0.012 & 0.030 & E/S0 \\
S3 & $21.22 \pm	 0.29$ & $7.03 \pm  0.67$ & 0.052 & $19.60 \pm	0.07$ & $5.21 \pm  0.28$ & 0.009 & 0.016 & S0   \\
T3 & $21.05 \pm	 0.20$ & $3.32 \pm  0.40$ & 0.035 & $20.30 \pm	0.08$ & $3.57 \pm  0.24$ & 0.022 & 0.034 & E/S0 \\
U3 & $20.38 \pm	 0.10$ & $4.09 \pm  0.13$ & 0.024 & $19.46 \pm	0.11$ & $4.06 \pm  0.06$ & 0.030 & 0.038 & E/S0 \\
A4 & $20.85 \pm	 0.02$ & $2.77 \pm  0.55$ & 0.048 & $19.75 \pm	0.13$ & $2.66 \pm  0.30$ & 0.036 & 0.007 & S0   \\
B4 & $20.43 \pm	 0.06$ & $2.36 \pm  0.15$ & 0.010 & $20.06 \pm	0.03$ & $2.72 \pm  0.08$ & 0.009 & 0.011 & E/S0 \\
D4 & $20.67 \pm	 0.05$ & $6.06 \pm  0.32$ & 0.025 & $19.73 \pm	0.07$ & $5.93 \pm  0.20$ & 0.017 & 0.011 & E/S0 \\
E4 & $20.13 \pm	 0.06$ & $2.68 \pm  0.23$ & 0.027 & $19.17 \pm	0.03$ & $2.56 \pm  0.12$ & 0.024 & 0.024 & E/S0 \\
H4 & $21.17 \pm	 0.10$ & $9.73 \pm  0.71$ & 0.080 & $20.25 \pm	0.10$ & $9.33 \pm  0.38$ & 0.054 & 0.043 & E    \\
I4 & $20.26 \pm	 0.38$ & $3.80 \pm  0.98$ & 0.081 & $19.43 \pm	0.24$ & $3.82 \pm  0.53$ & 0.054 & 0.051 & E/S0 \\
\hline
\end{tabular}
\end{table*}

\subsection{Error analysis}
\label{ssec:photoerr}

The main sources of error are given by: 
\begin{enumerate}
\item {\bf Fitting technique}. We estimate the error in terms of the 
half-difference of the values found with the two independent fitting
techniques used (luminosity profiles and 2Dl$\chi^2$ fitting).
\item {\bf PSF modelling}. We estimate this effect by fitting a
small galaxy in the corner of the chip ({\bf D3}) and a large one
in the centre of the chip ({\bf H4}) with a set of different PSFs.  We
use a Tiny Tim PSF generated on the galaxy position, a star present on
the chip, and a Tiny Tim PSF generated on the position of the
star. For {\bf D3}, which is observed on three frames with different
subpixel pointings, we produce a PSF by summing three subsampled PSFs
centred on the corresponding position within the pixel. The
r.m.s. scatter of the effective radius derived with the various PSFs
is 3 per cent for {\bf D3} and 2 per cent for {\bf H4} and the scatter of
$FP_{ph}$ is respectively 0.007 and 0.004. Therefore, this is a negligible
source of error for our purposes.
\item {\bf Sky subtraction and flat fielding}. We measured the contribution
to the error given by these uncertainties by varying the background by
1 per cent (which we take as upper limit to the errors in sky subtraction
and flat fielding) and by running the 2D fits. These errors are
smaller than the ones due to the fitting technique, but in some cases
they are not negligible.
\item {\bf Galactic extinction}. We adopt the error estimate by Schlegel et
al.\ (1998) for the reddening correction $\delta$E(B-V)=0.16E(B-V).
\item {\bf K-color correction}. For each galaxy, we computed the
colours of single-burst stellar population spectra (Bruzual \& Charlot
1993; GISSEL96 version) with ages between 0.5 and 20 Gyr, redshifted
to the galaxy redshift. We also computed K-color corrections for each
synthetic spectrum.  By interpolation we found the K-color correction
corresponding to the colour of each galaxy.  This procedure applied to
the colours shifted by a standard deviation provided upper and lower
limits to the K-color correction.  The error on the K-color correction
is the half-difference of the upper and lower
limits. Table~\ref{tab:Kcolor} lists the K-color corrections
($\Delta$m$_{B6}$ and $\Delta$m$_{V8}$) together with their errors
($\delta\Delta$m$_{B6}$ and $\delta\Delta$m$_{V8}$).

\end{enumerate}
The total error on the measured parameters listed in
Table~\ref{tab:photopar} is the quadratic sum of terms {\it i} and
{\it iii}. The uncertainties on the extinction correction and on the
K-color correction are added in quadrature to produce the total errors
listed in Table~\ref{tab:photorest}. The systematic uncertainty in the
zero point is not considered here and will be taken into account when
comparing the intermediate redshift results to the local ones in PIII.

\subsection{Classification}

\label{ssec:morph}

The objects are classified in a simplified scheme (E, E/S0, S0, S)
based on their morphology and luminosity profile. Among the early-type
systems, objects that are best-fitted by a pure \dv\, profile are
classified as E; objects for which the addition of an exponential
component gives a slightly better fit with no significant change in
the structural parameters are classified as E/S0; objects that are
significantly better described by an \dv plus exponential profile are
classified as S0. Objects that show a spiral pattern ({\bf B1, E3})
are classified as S. The classification is listed in
Table~\ref{tab:photorest}. Note that in this classification the usage
of the symbol E/S0 is different with the respect to the rest of the
paper, where it indicates the entire class of early-type galaxies.

\section{Spectroscopy}
\label{sec:spec}

A total of 13 dark nights at the ESO 3.6m telescope were awarded for
this spectroscopic project.  

The first pilot observing run (run 1, described in T99) took place in
1996, April 18-22.  Long slit ($ 1 \farcs 5 $ slit width) spectroscopy
was obtained with the EFOSC spectrograph, with the Orange 150 grism in
the spectral range between approximately 5200 and 7000 \AA\,
($512\times 512$ CCD, with pixel size $30 \mu m$ equivalent to about
$3.5 \tx{\AA} \times 0 \farcs 57 $).  The estimated seeing, measured with
R-band acquisition images, was $ 0
\farcs 8 $ FWHM.  

The following runs (run 2, 3, 4) took place in 1999, March 19-21,
August 11-14, and November 2-6. The imager-spectrograph EFOSC2 was
used with grism \# 9, which covers the range 4700-6700 \AA, with pixel
size of approximately 2\AA$\times 0\farcs32$ in $2\times2$ binning
mode. A narrower slit ($1''$) was used in order to achieve a better
resolution. The seeing ranged from $0\farcs8-\farcs2$ during run 2 and
3. Run 4 was subject to highly variable weather conditions: two nights
were lost for clouds and bad seeing ($2-3''$), one night had average
seeing conditions (0$\farcs 8-1''$), and during the last night we had
some episodes of excellent seeing (image quality 0$\farcs$6).

He-Ar lamps in the pointing position were taken before and after every
set of 2 exposures to provide accurate wavelength
calibration. Standard stars were observed at the beginning and at the
end of every night to provide relative flux calibration.

\begin{figure*}
\mbox{\epsfysize=18cm \epsfbox{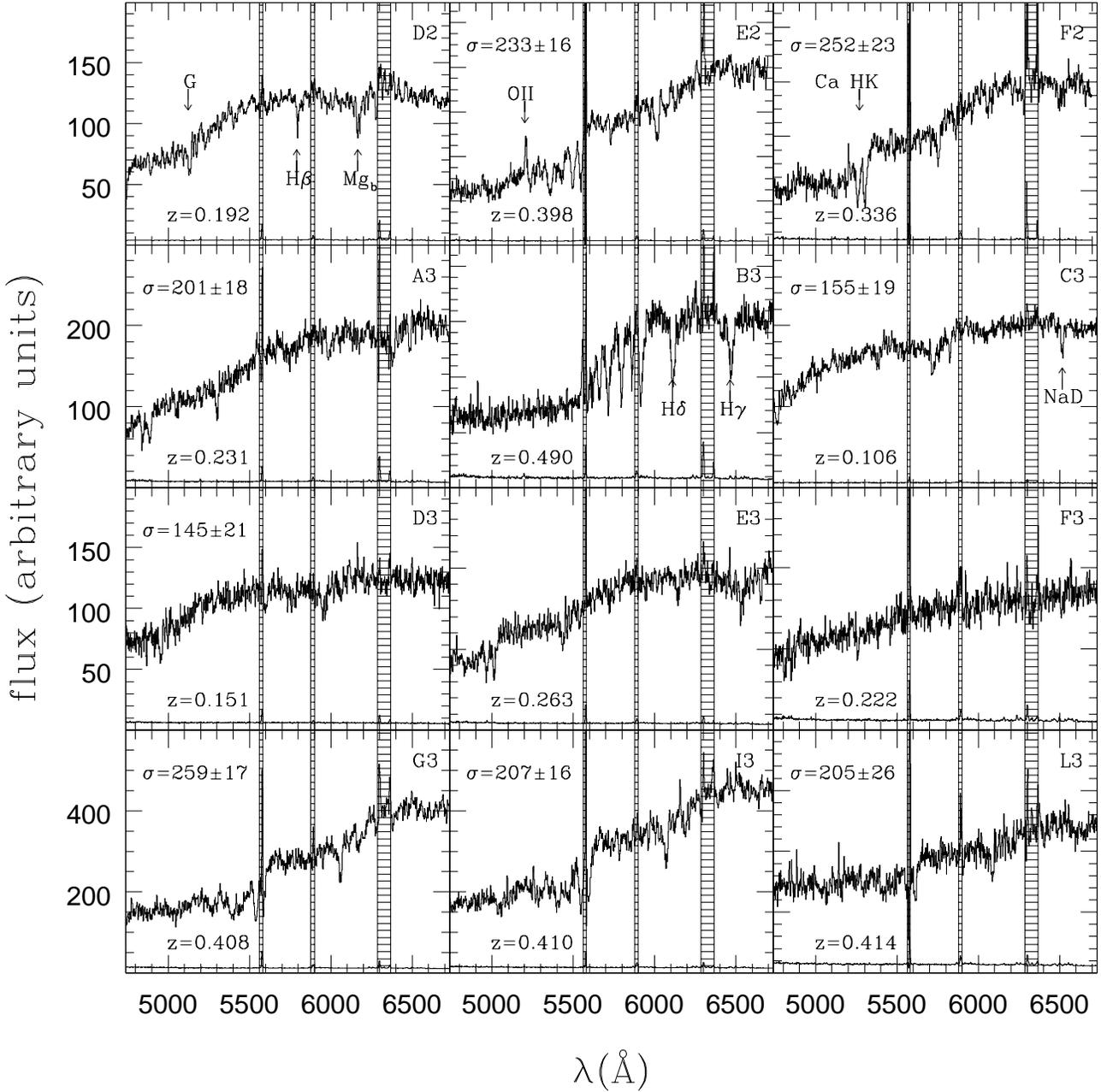}}
\caption{Wavelength calibrated spectra of the galaxies with average
S/N per pixel greater than 10 taken with EFOSC2, I. The shaded
spectral regions were affected by strong sky-emission lines. These
regions have not been used in the kinematic fit. The standard
deviation per pixel is also shown (lower curves). The Calcium doublet
(Ca H 3968.5 \AA\, and K 3933.7 \AA), the G band (4304.4 \AA),
H$\beta$, and Mg$_b$ (5175.4 \AA) are the most prominent absorption
features in this spectral region. Note the extremely strong Balmer
absorption features in the spectrum of {\bf B3} and the NaD (5892.5
\AA) line in the spectrum of the lowest redshift objects such as {\bf
C3}. Note [OII]3727 (\AA)in emission in the spectrum of {\bf E2}. The
central velocity dispersion in kms$^{-1}$ is shown in the upper left
corner when measured (see Section~4).}
\label{fig:spec1}
\end{figure*}

\begin{figure*}
\mbox{\epsfysize=18cm \epsfbox{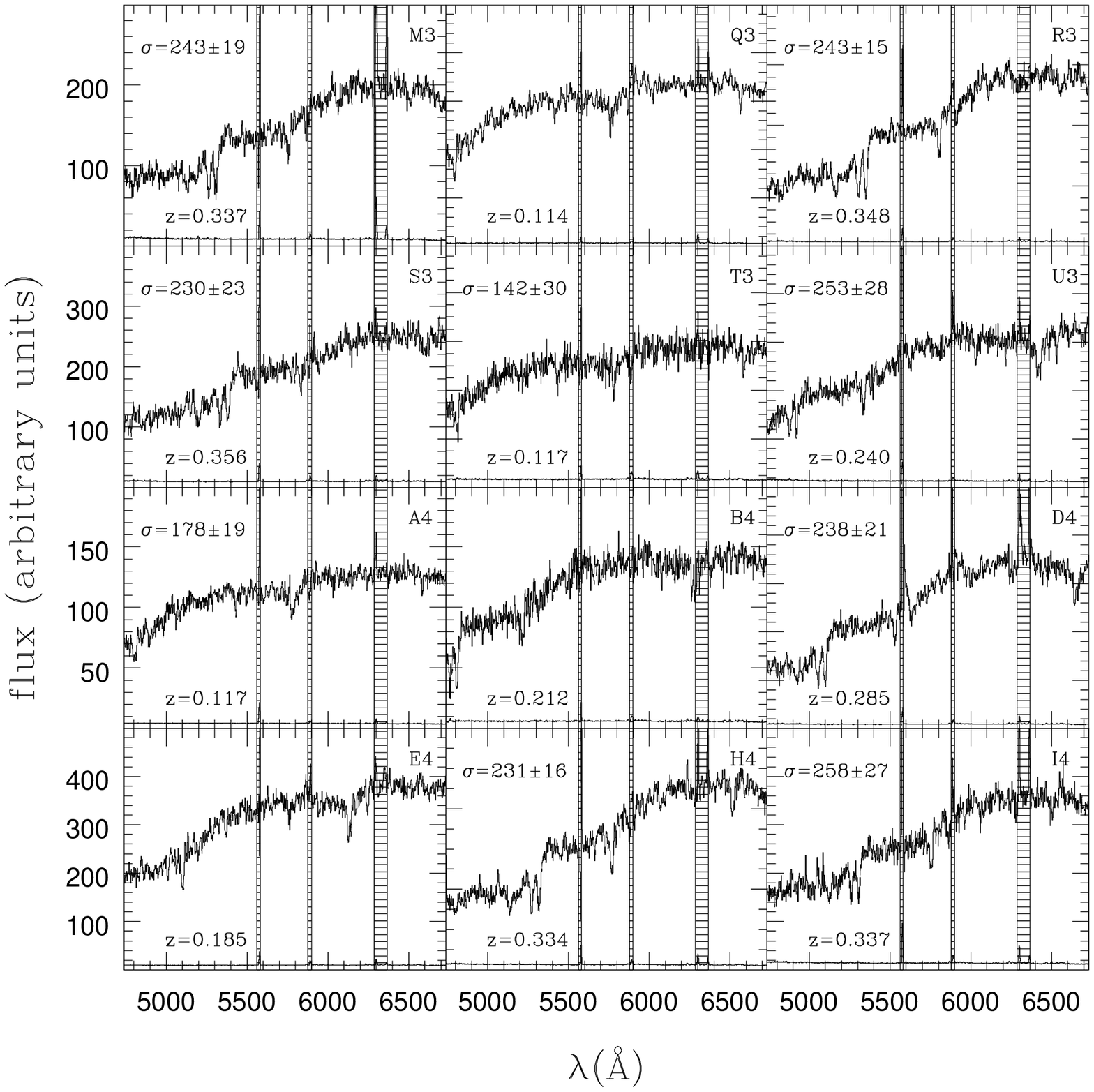}}
\caption{Wavelength calibrated spectra of the galaxies with
average S/N per pixel greater than 10 taken with EFOSC2, II. See
caption to Figure~\ref{fig:spec1}.}
\label{fig:spec2}
\end{figure*}

The spectra were reduced as described in T99 (see also Treu 2001). For
every galaxy, Table~\ref{tab:obj} gives total integration time, number
of exposures, S/N achieved.  In Figures~\ref{fig:spec1}
to~\ref{fig:spec2} the spectra with highest signal-to-noise collected
are shown. The run is identified by the digit in the name. The spectra
taken during run 1, shown in T99, are not repeated here.

\begin{figure}
\mbox{\epsfxsize=8cm \epsfbox{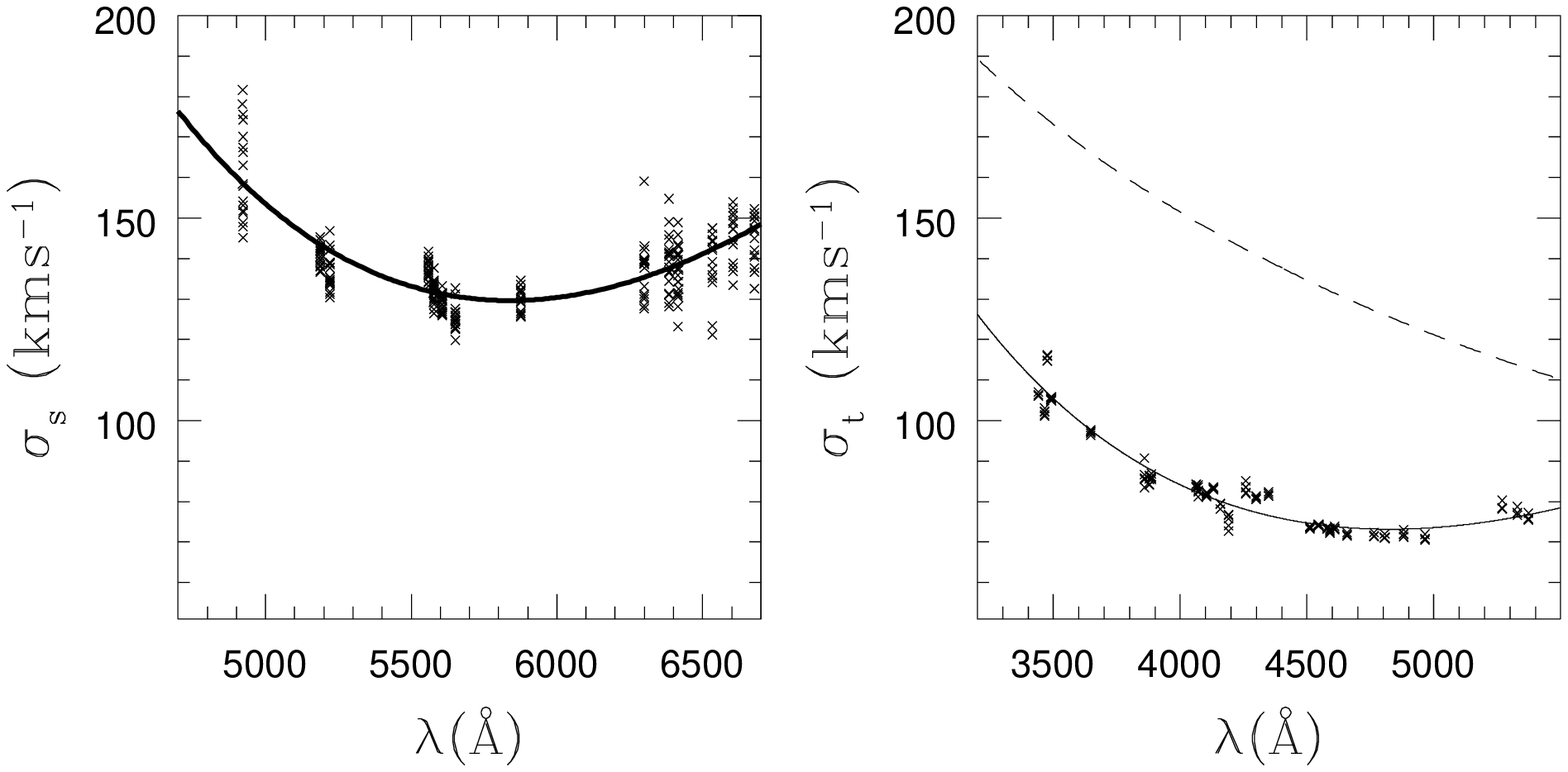}}
\caption{Left: resolution ($\sigma_s$) as a function of wavelength as
determined from the width of sky and He-Ar lines for the EFOSC2
setup. Right: resolution ($\sigma_t$) as a function of wavelength for
the SK stellar template library (the points represent the measured
resolution, the solid line is the best fitting parabola; see Table
\ref{tab:templates} and Section~\ref{ssec:templates}); the resolution
of the JHC library (4.76\AA\, see Section~\ref{ssec:templates}) is
shown for comparison as a dashed line.}
\label{fig:newres}
\end{figure}

For runs 2, 3, and 4, by taking advantage of the higher number of
spectra available with the same instrumental setup, we were able to
determine the resolution of our final combined spectra more accurately
than in run 1 (see T99). We selected a set of 3 sky lines and 12 He-Ar
lines that are unblended at our resolution and sufficiently strong and
isolated and we measured their FWHM on each combined sky and He-Ar
calibration spectrum. The results were very stable from frame to frame
and from sky to He-Ar lines, so we fitted the entire set of
measurements as a function of wavelength. The measurements of the
resolution ($\sigma_s$ in \kms) are well fitted by a parabola, as
shown in Figure~\ref{fig:newres}, left panel.
\subsection{Fitting kinematic parameters}

The internal velocity dispersion is derived by comparison of the
observed galaxy spectrum to a stellar spectrum taken with the same
resolution under the assumption that the galaxy spectrum can be well
described by the sum of the spectra emitted by stars with a Gaussian
velocity distribution.

In the first exploratory study (T99) we derived kinematic information
for our sample of galaxies with two different and independent methods:
{\it i)} the {\sc Gauss-Hermite Fourier Fitting Software} (Franx,
Illingworth \& Heckman 1989; van der Marel \& Franx 1993), hereafter
GHFF, and {\it ii)} the {\sc Fourier Quotient} (Sargent et al.\ 1977)
in the version modified by Rose (see Dressler 1979) and by Stiavelli,
M{\o}ller \& Zeilinger (1993). The results of the two methods were
mutually consistent, within the estimated errors. However, the GHFF
allows for a reliable estimate of the errors, is less sensitive to
template mismatches (van der Marel \& Franx 1993), and provides
insight in the fitting procedure. Finally, it provides a quality
parameter (the $\chi^2$) and the residual spectrum. Therefore, here we
will use only the GHFF software.

\subsection{Stellar templates}
\label{ssec:templates}

A good library of stellar spectra is a key ingredient to measure the
internal kinematics of a galaxy. It has to cover the largest possible
range of spectral types so that stellar population effects can be
studied. In addition, the spectral resolution of the stellar template
($\sigma_t$) must be better than the instrumental resolution of the
setup used for galactic spectroscopy. The stellar library used in T99
(Jacoby, Hunter \& Christian 1984; hereafter the JHC library), given
its resolution ($\sim$ 4.5 \AA\, FWHM, i.e. $\sigma_t=133$ \kms at
4300 \AA), can only be used out to $z\sim0.2$ with our spectroscopic
setup.

David Soderblom and Jeremy King kindly provided us with a library of
stellar spectra at higher resolution for the highest redshift part of
the sample (hereafter the SK library).  The stars, listed in
Table~\ref{tab:templates}, were observed at the Kitt Peak National
Observatory Coud\'e Feed Telescope, with the Coud\'e CCD Spectrograph,
on May 22 1999. The stars were chosen according to observability from
a list of giant stars covering all spectral types from A0 to K7. In
addition, a few dwarf star spectra were also taken. The spectral range
covered is 3120--5510 \AA. The resolution of the SK library, as
measured from calibration lamps, is shown in Figure~\ref{fig:newres},
right panel.
\begin{table}
\caption{Stars observed as stellar templates (SK library; see 
Section~\ref{ssec:templates} for details). For each star we list the
spectral type, name of the star, magnitude and colour, taken.}
\begin{tabular}{lccc}
type & star & V & B-V \\
\hline 
A2III & HR4343 & 4.48 & 0.03 \\
A9III & HR4584 & 6.50 & 0.22 \\
F3III & HR5783 & 6.45 & -    \\
F5III & HR4191 & 5.18 & 0.33 \\
F9III & HR4451 & 6.05 & 0.60 \\
G4III & HR4255 & 5.66 & 0.83 \\
G4III & HR4558 & 5.30 & 0.88 \\ 
G5III & HR3922 & 5.93 & 0.89 \\
G8III & HR5888 & 5.23 & 1.02 \\ 
G8III & HR6770 & 4.64 & 0.96 \\
G9III & HR4609 & 5.80 & 1.01 \\
K0III & HR4287 & 4.08 & 1.09 \\
K2III & HR6299 & 3.20 & 1.15 \\
K3III & HR4365 & 5.32 & 1.20 \\
K3III & HR4521 & 5.27 & 1.27 \\
K6III & HR4672 & 5.81 & 1.30 \\ 
K7III & HR6159 & 4.84 & 1.49 \\
F5V   & HR4657 & 6.11 & 0.46 \\
F9V   & HR4540 & 3.61 & 0.55 \\
G0V   & HR4845 & 5.95 & 0.55 \\
\hline
\end{tabular}
\label{tab:templates}
\end{table}

Unfortunately, the limited red coverage of the spectra in the SK
library did not allow us to use them for the low redshift objects of
the sample ($z<0.21$).  Therefore, we used the JHC library for the low
redshift  part of the sample ($z<0.21$) and the SK library for the high
redshift part of the sample ($z>0.21$). 

In order to derive an accurate and consistent value for the resolution
of the JHC library, we measured it using the SK library. We divided
the overlapping region (3510-5510 \AA) into four intervals of 500
\AA\, each and we fitted the ``velocity dispersion'' of each interval
of the JHC spectra using the SK template that more closely matches the
spectral type.  By adding in quadrature the resolution of the SK
library, we found the intrinsic resolution of the JHC library
($4.76\pm0.30$ \AA\, FWHM), in agreement with the value given by Jacoby
et al.\ (1984). We used this value as intrinsic resolution of the
JHC library.

The stellar templates were redshifted to the redshift of each object
and then broadened with a Gaussian characterized by a width equal to
the quadratic difference between the instrumental resolution
($\sigma_s$) and the resolution of the template library ($\sigma_t$).

\subsection{Results and error discussion}

\label{ssec:kineres}

As in T99, we fitted each galaxy with each of the stellar templates
available\footnote{The spectra from T99 were fitted again using the SK
library, except for galaxy {\bf A1} that was fitted with the JHC
library spectra at the new slightly different resolution.}.  This
procedure allows us to estimate the uncertainty given by template
mismatches. As noted in T99, the late G to early K giant stars
provided the best $\chi^2$, the smallest residual, and the most stable
results. Cooler stars produced systematically higher velocity
dispersions and hotter stars systematically lower ones.

As best estimate of the velocity dispersion and redshift (\sgh, \zgh\,
in Table~\ref{tab:kineres}), we adopted the average of the values
obtained with templates in the spectral range G4-K0III. In order to
achieve homogeneous results with the two libraries, we used the same
number of templates and closely matching spectral type distributions:
2xG4, 1xG5, 2xG8, 1xG9, 1xK0 for SK and 1xG5, 2xG6, 1xG7, 1xG8, 1xG9,
1xK0 for JHC. In Table~\ref{tab:kineres} we also list the best-fitting
template for each galaxy. Note that the best-fitting template spans
the entire range of spectral types. Two estimates of the errors are
also given, as in T99: the random component, which is the fit formal
uncertainty obtained for the best-fitting template ($\delta$); the
systematic component ($\Delta$), resulting from template mismatches,
computed as the r.m.s. scatter of the values obtained with different
templates.

\begin{table*}

\caption{Kinematic results from the GHFF fit. For each galaxy we list
the average velocity dispersion (\sgh) together with the random error
($\delta$\sgh) and the systematic error ($\Delta$\sgh).  The random
error is the fit formal uncertainty for the best-fitting template,
while $\Delta$\sgh\, is the scatter of the values obtained with the
templates in the range G4-K0 III (see Section~\ref{ssec:kineres}). We
also list the redshift (\zgh) with its errors, the spectral type of
the best-fitting template (Type), the signal-to-noise ratio (S/N), and
the matching to the relevant selection criteria (sel; see
Section~\ref{ssec:sample}). Only galaxies with $\tx{S/N}>12$ are
listed (see Section~\ref{ssec:kineres}). The values of velocity
dispersion found for {\bf B1, D2, E3, Q3, B4, E4} are too small to be
acceptable (see discussion in Section~\ref{ssec:kineres}): these
objects have been excluded from the table.}
 
\label{tab:kineres}
\begin{tabular}{lccccccccc}

gal & $\sigma_{\tx{ghff}}$ & $\delta\sigma_{\tx{ghff}}$ & $\Delta$\sgh
& \zgh & $\delta$ \zgh & $\Delta$ \zgh & Type & S/N & sel\\

\hline
A1 & 227 & 16 & 14 & 0.1466 & 4.3$\cdot 10^{-5}$ & 1.2$\cdot 10^{-4}$ &  G8 & 65 & 2 \\
D1 & 290 & 21 & 12 & 0.3848 & 6.8$\cdot 10^{-5}$ & 1.4$\cdot 10^{-4}$ &  G4 & 22 & 2 \\
E1 & 198 & 25 &  7 & 0.2940 & 8.4$\cdot 10^{-5}$ & 1.3$\cdot 10^{-4}$ &  K0 & 17 & 2 \\
G1 & 220 & 17 & 10 & 0.2946 & 5.9$\cdot 10^{-5}$ & 1.3$\cdot 10^{-4}$ &  G8 & 35 & 2 \\
I1 & 190 & 20 &  4 & 0.2929 & 6.5$\cdot 10^{-5}$ & 1.3$\cdot 10^{-4}$ &  G5 & 22 & 2 \\
E2 & 219 & 12 &  3 & 0.3977 & 4.4$\cdot 10^{-5}$ & 1.4$\cdot 10^{-4}$ &  G4 & 24 & 2 \\
F2 & 237 & 14 & 14 & 0.3364 & 5.3$\cdot 10^{-5}$ & 1.3$\cdot 10^{-4}$ &  G9 & 18 & 2 \\
A3 & 189 & 15 &  4 & 0.2311 & 4.9$\cdot 10^{-5}$ & 1.2$\cdot 10^{-4}$ &  G5 & 19 & 2 \\
C3 & 146 & 15 &  9 & 0.1057 & 3.8$\cdot 10^{-5}$ & 1.4$\cdot 10^{-4}$ &  G8 & 27 & 2 \\
D3 & 136 & 18 &  8 & 0.1511 & 4.8$\cdot 10^{-5}$ & 1.0$\cdot 10^{-4}$ &  G9 & 17 & 0 \\
G3 & 244 & 11 &  7 & 0.4081 & 4.5$\cdot 10^{-5}$ & 1.4$\cdot 10^{-4}$ &  G5 & 21 & 2 \\
I3 & 194 & 11 &  8 & 0.4103 & 4.1$\cdot 10^{-5}$ & 1.4$\cdot 10^{-4}$ &  G5 & 22 & 2 \\
L3 & 193 & 18 & 15 & 0.4141 & 6.8$\cdot 10^{-5}$ & 1.4$\cdot 10^{-4}$ &  G5 & 13 & 1 \\
M3 & 228 & 15 &  5 & 0.3374 & 5.4$\cdot 10^{-5}$ & 1.4$\cdot 10^{-4}$ &  G4 & 16 & 2 \\
R3 & 228 & 11 &  5 & 0.3482 & 4.0$\cdot 10^{-5}$ & 1.3$\cdot 10^{-4}$ &  G5 & 24 & 2 \\
S3 & 216 & 19 &  5 & 0.3559 & 7.2$\cdot 10^{-5}$ & 1.1$\cdot 10^{-4}$ &  G4 & 18 & 2 \\
T3 & 133 & 23 & 16 & 0.1174 & 5.2$\cdot 10^{-5}$ & 1.0$\cdot 10^{-4}$ &  G6 & 14 & 2 \\
U3 & 238 & 14 & 21 & 0.2401 & 5.2$\cdot 10^{-5}$ & 1.2$\cdot 10^{-4}$ &  G9 & 18 & 0 \\
A4 & 167 & 15 &  8 & 0.1171 & 4.1$\cdot 10^{-5}$ & 1.2$\cdot 10^{-4}$ &  G7 & 25 & 2 \\
D4 & 224 & 15 & 11 & 0.2848 & 5.1$\cdot 10^{-5}$ & 1.2$\cdot 10^{-4}$ &  G4 & 32 & 2 \\
H4 & 217 & 12 &  5 & 0.3399 & 4.4$\cdot 10^{-5}$ & 1.4$\cdot 10^{-4}$ &  G9 & 25 & 2 \\
I4 & 243 & 20 & 13 & 0.3369 & 7.2$\cdot 10^{-5}$ & 1.3$\cdot 10^{-4}$ &  G4 & 19 & 2 \\
\hline
\end{tabular}
\end{table*}

It is intuitive and well known (e.g. Bender 1990; J{\o}rgensen, Franx
\& Kj{\ae}rgaard 1995; T99) that the S/N needed to measure velocity
dispersions properly scales with the ratio between the instrumental
resolution and the velocity dispersion itself. In other words, when
the velocity dispersion is smaller than the instrumental resolution a
higher minimum signal-to-noise is needed with respect to the case when
the velocity dispersion is larger than the instrumental resolution.

\begin{figure}
\mbox{\epsfysize=8cm \epsfbox{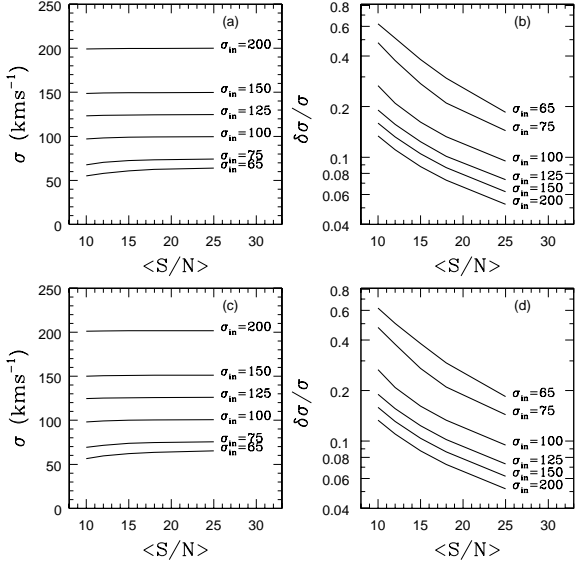}} 
\caption{Results of Montecarlo simulations for the EFOSC2 setup. For a
range of values of velocity dispersion ($\sigma_{\tx{in}}$) the
velocity dispersion recovered is plotted as a function of S/N (using
the same star as ``galaxy'' and template in panel a, and using a
different star as template in panel c). The scatter of the
distribution is plotted in panels b (same star as template and galaxy)
and d (different star as template and galaxy). At $\tx{S/N}<15$
velocity dispersions below 100 \kms\, are systematically
underestimated. Moreover, for small velocity dispersions, the
uncertainty is large even for higher values of S/N.}
\label{fig:restest}
\end{figure}

In practice, for any given S/N and resolution, there is a lower limit
on the velocity dispersion measurable without introducing significant
bias (for example, Bender 1990 considers half of the instrumental
resolution to be the minimum measurable value). Quantitatively, these
numbers depend on the instrumental setup (resolution, sampling).
Therefore, we addressed the problem by numerical simulation of our
measurement procedure. For a range of values of velocity dispersions
($\sigma_{\tx{in}}=65$, 75,~100,~125,~150,~175,~200 \kms) and of S/N
(10,~12,~15,~18,~25), we created 1000 toy galaxy spectra with
artificial noise and our instrumental sampling and resolution. We then
recovered the velocity dispersion with the GHFF software.  The results
are plotted in Figure~\ref{fig:restest}.

For each input value of $\sigma_{\tx{in}}$ we plot the average
recovered $\sigma$ as a function of S/N, using the same stellar
spectrum as ``galaxy'' and template (G8 III; panel a) or using a
different star as template (G9 III; panel c). It is noticed that the
smallest values of $\sigma_{\tx{in}}$ are systematically
underestimated by the fit for low values of S/N (by as much as 13 per
cent for $\sigma_{\tx{in}}=65$~\kms\, and $\tx{S/N}=10$) but for
$\sigma_{\tx{in}}>100$ \kms\, the effect becomes negligible ($<2$ per
cent at $\tx{S/N}=10$, $<1$ per cent at $\tx{S/N}=12$ for
$\sigma_{\tx{in}}=100$ \kms\, and using a different template). If --
instead of the average of $\sigma$ -- the average of the logarithm of
$\sigma$ weighted on the formal variance is used as estimator
(J{\o}rgensen et al.\ 1995), the effect is reversed: the velocity
dispersion is now overestimated for small values of $\sigma_{\tx{in}}$
(by up to 15 per cent at $\sigma_{\tx{in}}=65$ \kms and
$\tx{S/N}=10$). In addition, high values of $\sigma_{\tx{in}}$ (from
200~\kms\, in our simulation) become underestimated by 2 per cent at
$\tx{S/N}=10$. The conclusion is that the systematic uncertainty on
small values of S/N is hard to correct, so that it is suggested that
these data should be rejected.  In our particular case we will only
accept for measurement spectra with $\tx{S/N}>12$ (thus rejecting {\bf
F1, B3, F3}). In Figure~\ref{fig:restest} we also plot the relative
scatter around the mean for the same configurations (template G8 III,
panel b; template G9 III, panel d). For a given S/N the uncertainty on
the velocity dispersion increases dramatically for the smallest values
of $\sigma_{\tx{in}}$, while it tends to become constant for velocity
dispersions much larger than the resolution. This is consistent with
what we find for galaxies {\bf B4} and {\bf E3} (\sgh~$=64$ and 98
\kms): despite the good S/N of the spectrum the formal uncertainty on
$\sigma_{\tx{in}}$ is $\sim 35$ and $\sim 18$ per cent. We do not
consider these values to be successfully measured. Similarly, the
small value of velocity dispersion of galaxy {\bf B1} was rejected in
T99.

Since most of the objects at $z<0.21$ have high S/N spectra, the
problems discussed above are negligible. Nevertheless, the uncertainty
in the JHC spectral library resolution ($\sim120 \pm 8$ \kms\, at 5000
\AA) can cause systematic effects on the low velocity dispersion
objects. We ran extensive simulations (as above) to measure this
effect. For velocity dispersions smaller than the resolution of the
library, the uncertainty is important (for example $\sim 10$ per cent at 100
\kms\, with $\tx{S/N}=25$), but it drops quickly with increasing velocity
dispersion (6 per cent at 125 \kms\, with $\tx{S/N}=25$; 4 per cent at 150
\kms). In practice, at $\sigma_{\tx{in}}=125$ \kms\, this effect is already
smaller than the typical uncertainty. We will not consider galaxies
{\bf D2, Q3, E4} as successfully measured because their velocity
dispersion is smaller than this limit.

The uncertainty in the measurement of the instrumental resolution of
EFOSC ($\Delta \lambda = 8.39\pm 0.26$\AA) introduces a systematic
error in the measurement of the velocity dispersion that is negligible
in our case.  For example, let us consider {\bf A1}, the galaxy with
the highest signal-to-noise spectrum. The random error is estimated to
be 6 per cent, the systematic error due to template mismatches is 7
per cent; therefore, the additional 3 per cent associated with
the uncertainty in the measurement of instrumental resolution, to be
added in quadrature, does not change the total error significantly.
As a check, we repeated the fits with templates broadened to the
resolution changed by a standard deviation. For galaxy {\bf I1} we
obtained the largest difference (7 \kms), still negligible with
respect to the other sources of error. The effect on the uncertainty
on the resolution of EFOSC2 is smaller and totally negligible.

\begin{figure}
\mbox{\epsfysize=8cm \epsfbox{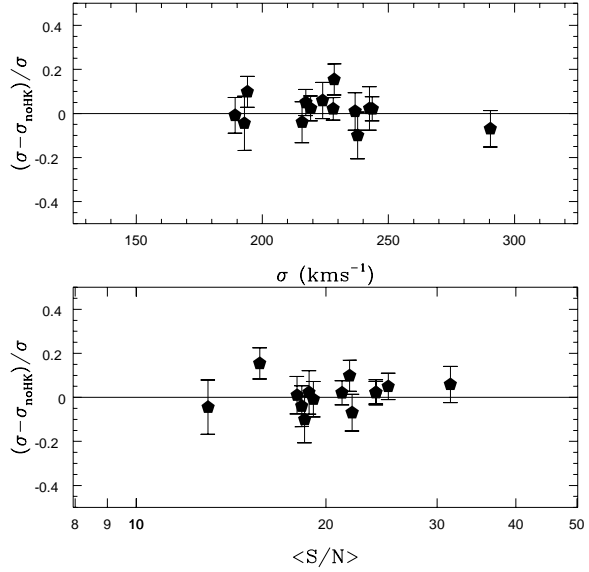}} 
\caption{Relative difference in velocity dispersion as measured with
and without Ca H and K as a function of velocity dispersion (upper
panel) and average S/N per pixel (lower panel). The error bars are the
quadratic sum of the systematic and random errors, as defined in the
Section~\ref{ssec:kineres}. The average difference is 1.4 per cent with a
scatter of 6.6 per cent. The average difference is negligible with respect
to the errors (see Table~\ref{tab:kineres}) and consistent with the
scatter found.}
\label{fig:CaHK}
\end{figure}

The spectra of galaxies {\bf D1, E2, F2, G3, I3, L3, M3, R3, S3, U3,
D4, H4,} and {\bf I4} include the lines Ca H and K. As noted in T99,
even though widely used to perform such measurements (Dressler 1979),
these lines have been reported to induce a slight overestimate of the
velocity dispersion (Kormendy 1982; see also Kormendy \& Illingworth
1982). Kormendy \& Illingworth (1982) suggest that the problem may be
caused by the intrinsic width of the lines or by the steepness of the
continuum in that spectral region. However, Dressler (1984) suggests
that they may be best suited for the measurement of large velocity
dispersions for faint objects, such as {\bf D1}.  For these reasons,
we performed the kinematic fit also by masking the region of Ca H and
K. In Figure~\ref{fig:CaHK} the variation of $\sigma$ as a function of
velocity dispersion and S/N is plotted. The velocity dispersion
obtained with Ca H and K, as listed in Table~\ref{tab:kineres}, is on
average 1.4 per cent higher than the one obtained by masking this
region, with a 6.6 per cent scatter. The effect is within the
estimated error and the scatter is likely to be induced by the loss of
information caused by masking a significant part of the spectrum. We
conclude that with this kind of resolution, sampling, signal-to-noise
ratio, and treatment of the continuum, the presence of Ca H and K does
not alter significantly the measurement of velocity dispersion.

\begin{figure}
\mbox{\epsfysize=8cm \epsfbox{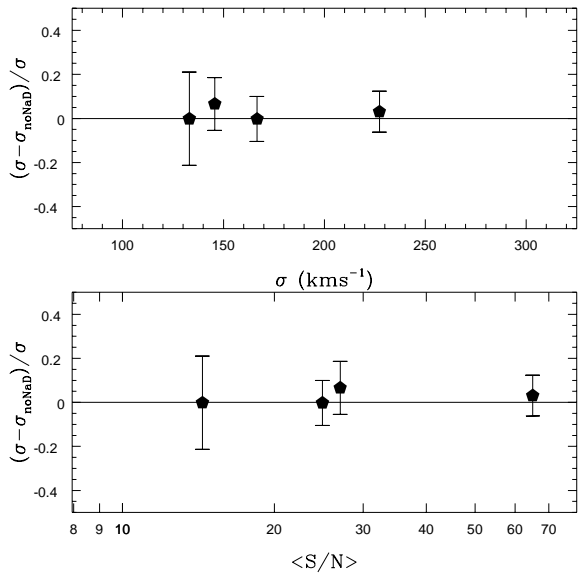}} 
\caption{Relative difference in velocity dispersion as measured with
and without NaD as a function of velocity dispersion (upper panel) and
average S/N per pixel (lower panel). The error bars are the quadratic
sum of the systematic and random errors, as defined in the
Section~\ref{ssec:kineres}. The average difference is 2.3 per cent,
negligible with respect to the errors (see Table~\ref{tab:kineres}).}
\label{fig:NaD}
\end{figure}

Similarly, the spectra of {\bf A1, C3, T3}, and {\bf A4} include the
region of NaD, which may be affected by interstellar absorption even
in elliptical galaxies (e.g. Dressler 1984). We repeated the analysis
by excluding the NaD region (see Figure~\ref{fig:NaD}) and found an
average velocity dispersion higher by 2.3 per cent. As above, we
conclude that the presence of NaD in the spectrum does not alter
significantly the result, and used in Table~\ref{tab:kineres} the
values found by using the entire spectral range.

The total error on the redshift, taking into account the error on
wavelength calibration, is less than 0.0005, when the GHFF fit was
performed and less than 0.002 in the cases where the redshift has been
measured by identifying the main spectral features.

\subsection{Aperture correction}

\label{ssec:apcorr}

Spatially resolved kinematics measurements in nearby galaxies (e.g.,
Capaccioli et al.\ 1993; Carollo \& Danziger 1994a,b; Bertin et al.\
1994) have shown that the velocity dispersion varies (generally it
declines) with radius. For this reason, central velocity dispersion is
generally different from the velocity dispersion measured from the
spectrum integrated over the entire galaxy, as is generally available
at intermediate redshift. A correction is thus required.  As in T99,
we model the velocity dispersion profile with a power law,
\begin{equation} 
\sigma(r)\propto \left(\frac{r}{r_e}\right)^d,
\end{equation} 
with $-0.1<d<0$.
The desired correction can be computed numerically from 
\begin{equation} 
\sigma^2({\mathcal A})\simeq\int_{\mathcal{A}} 2 \pi r dr \sigma^2(r) DV(r). 
\end{equation}
Here we have assumed that the value obtained by measuring $\sigma^2$
within an aperture ${\mathcal A}$ is the average of $\sigma^2(r)$
weighted by the luminosity density, modelled as an \dv law
appropriately normalized and indicated by $DV(r)$.  With respect to
T99, the way we compute the correction has been improved by taking
into account the effect of seeing. The effect of seeing is to smear
the dependence of the correction on $r_{\tx{e}}$, whenever the two
quantities are comparable. We computed the correcting factor
${\mathcal B}(d)=\sigma/$\sgh\, to an equivalent aperture of radius
$r_{\tx{e}}/8$ (see J{\o}rgensen et al.\ 1995; T99) for a range of
values of seeing ($0\farcs8-1\farcs2$), effective radius (0$\farcs
5-2''$), and number of lines used. The average correcting factor,
computed as ${\mathcal B}\equiv[{\mathcal B}(-0.1)+{\mathcal
B}(0)]/2$, ranges from 1.055 to 1.075. The uncertainty related to the
intrinsic variety of velocity dispersion profiles (modelled by $d$) is
larger than this interval. Therefore, we adopt the mean correcting
factor ${\mathcal B}=1.065\pm0.037$, where the error is estimated as
${\mathcal B}\equiv[{\mathcal B}(-0.1)-{\mathcal B}(0)]/2\sqrt{3}$ as
in T99.  The values obtained for the central velocity dispersion are
shown in Figures~\ref{fig:spec1} and~\ref{fig:spec2} together with the
spectrum of each galaxy.

%
%
\subsection{Relative flux calibration}

The instrumental response was measured at the beginning and at the end
of every night, by observing spectrophotometric standard stars at
parallactic angle through a $5''$ slit. The response was stable from
night to night within 2-3 per cent over the entire spectral range when the
same star was used for calibration. Different stars provided similar
responses within 5~ per cent.  We used the average response function to
correct the spectra. We estimate the relative flux calibration
uncertainty to be $\sim 5$ per cent, dominated by the systematic differences
found when different stars are used.

\section{Summary}
\label{sec:sampdisc}

In this paper we have presented the surface photometry and
spectroscopy of a sample of HST selected early-type galaxies. The
sample is chosen mainly based on morphology and colour. As described
in Section~\ref{sec:sample}, our sample is a fair sample of the
early-type galaxies found in the Medium Deep Survey, with respect to
morphological selection, colour distribution, bulge-to-total
luminosity distribution, and luminosity distribution. As already
noticed by other authors (see, e.g., Totani \& Yoshii 1998), at any
given redshift early-type galaxies span a significant range of
colours, from the red envelope of old cluster ellipticals bluewards
(Figure~\ref{fig:colorz}).

\begin{figure}
\mbox{\epsfysize=8cm \epsfbox{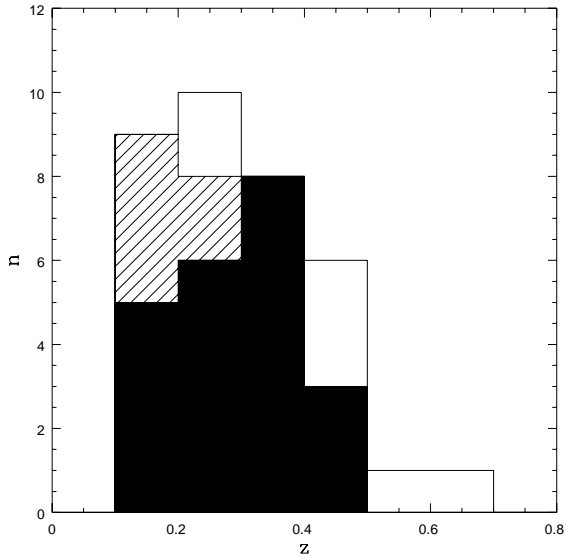}} 
\caption{Distribution of redshift for the sample galaxies (empty histogram). 
The galaxies with measurable velocity dispersion ($\tx{S/N}>12$) are
plotted as hatched histogram. The velocity dispersions smaller than
our resolution limit were rejected based on Montecarlo simulations
(see Section~\ref{ssec:kineres}). The galaxies with measured velocity
dispersion are plotted as filled histogram.}
\label{fig:histoz}
\end{figure}

In Section \ref{sec:photo} we have presented the surface photometry of 35
objects in the redshift range $z=0.11-0.65$ (see
Figure~\ref{fig:histoz}). The photometric structural parameters
(effective radius and effective surface brightness) have been derived
with two different techniques: isophote profile fitting (\dv and \dv +
exponential disc) and 2 dimensional fitting (\dv only). The value of
the structural parameters changes significantly (r.m.s scatter of
15 per cent in $r_{\tx{e}}$) with the different modelling, but the
combination of effective radius and surface brightness that enters the
FP is remarkably stable (r.m.s. scatter 0.03).

In Section \ref{sec:spec} we have described spectroscopic measurements
obtained at the ESO-3.6m telescope. Redshifts are obtained for all the
35 objects with measured surface photometry.  Using Montecarlo
simulations, we have studied extensively the systematic errors in the
measurement of $\sigma$, due to instrumental resolution and finite
S/N. The systematic errors are found to be significant for values of
$\sigma$ comparable to the instrumental resolution and for small
S/N. Based on these simulations, we measure velocity dispersion only on
the 28 spectra with average S/N per pixel larger than 12, and we
reject all values of $\sigma<125$~\kms. In the end, we have obtained robust
determination of velocity dispersion for 22 galaxies.

A range of stellar templates is used to recover the internal velocity
dispersion; the best-fitting template is found to span the entire
range G4-K0 III. In this way, we are able to investigate the
systematic errors induced by the mismatch between the galactic stellar
population and the stellar template used. The average error on
internal velocity dispersion is found to be 8 per cent (plus 5 per cent
systematic).  No significant offset in the average velocity dispersion
is found when Ca H and K and NaD lines are masked in the kinematic
fit.

\section{Acknowledgments}

This work is based on observations collected at the European Southern
Observatory (La Silla) under programmes 62.O-0592, 63.O-0468, and
64.O-0281 and with the NASA/ESA Hubble Space Telescope, obtained at
the Space Telescope Science Institute, which is operated by
Association of Universities for Research in Astronomy, Inc.\ (AURA),
under NASA contract NAS5-26555. Tommaso Treu was financially supported
by the Space Telescope Science Institute Director Discretionary
Research Fund grant 82228 and by the Italian Ministero
dell'Universit\`a e della Ricerca Scientifica e Tecnologica.  The use
of the Gauss-Hermite Fourier Fitting Software developed by R.~P.~van
der Marel and M.~Franx is gratefully acknowledged. We are grateful to
David Soderblom and Jeremy King for providing us with the library of
stellar templates used in the kinematic measurement. We thank
R.~J. Smith for his comments that improved the presentation of the
results.

\appendix

\section{Images of noteworthy galaxies}
\label{app:2D}

In this appendix we show the images of the two noteworthy galaxies
{\bf E3} and {\bf F3}.  Galaxy {\bf E3} shows a clear spiral pattern
in the residuals. The center of galaxy {\bf F3} is obscured by a dust
lane; the luminosity profile has been measure by correcting the dust
extinction as described in Section~\ref{sec:photo}. The images of all
the galaxies are available at the MDS world wide web site at URL
http://archive.stsci.edu/mds/.

\begin{figure*}
\mbox{
\mbox{\epsfysize=4cm \epsfbox{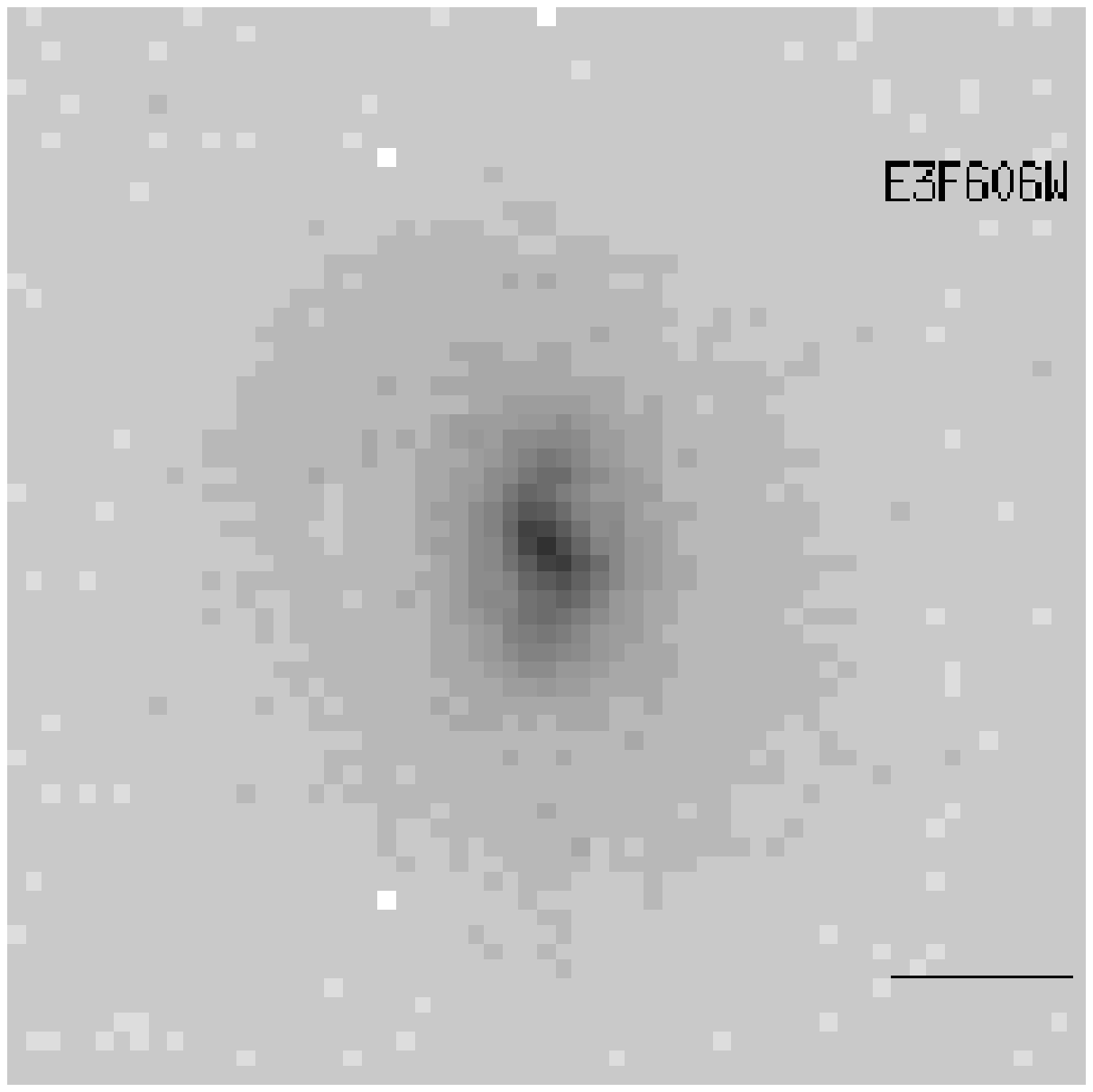}}
\mbox{\epsfysize=4cm \epsfbox{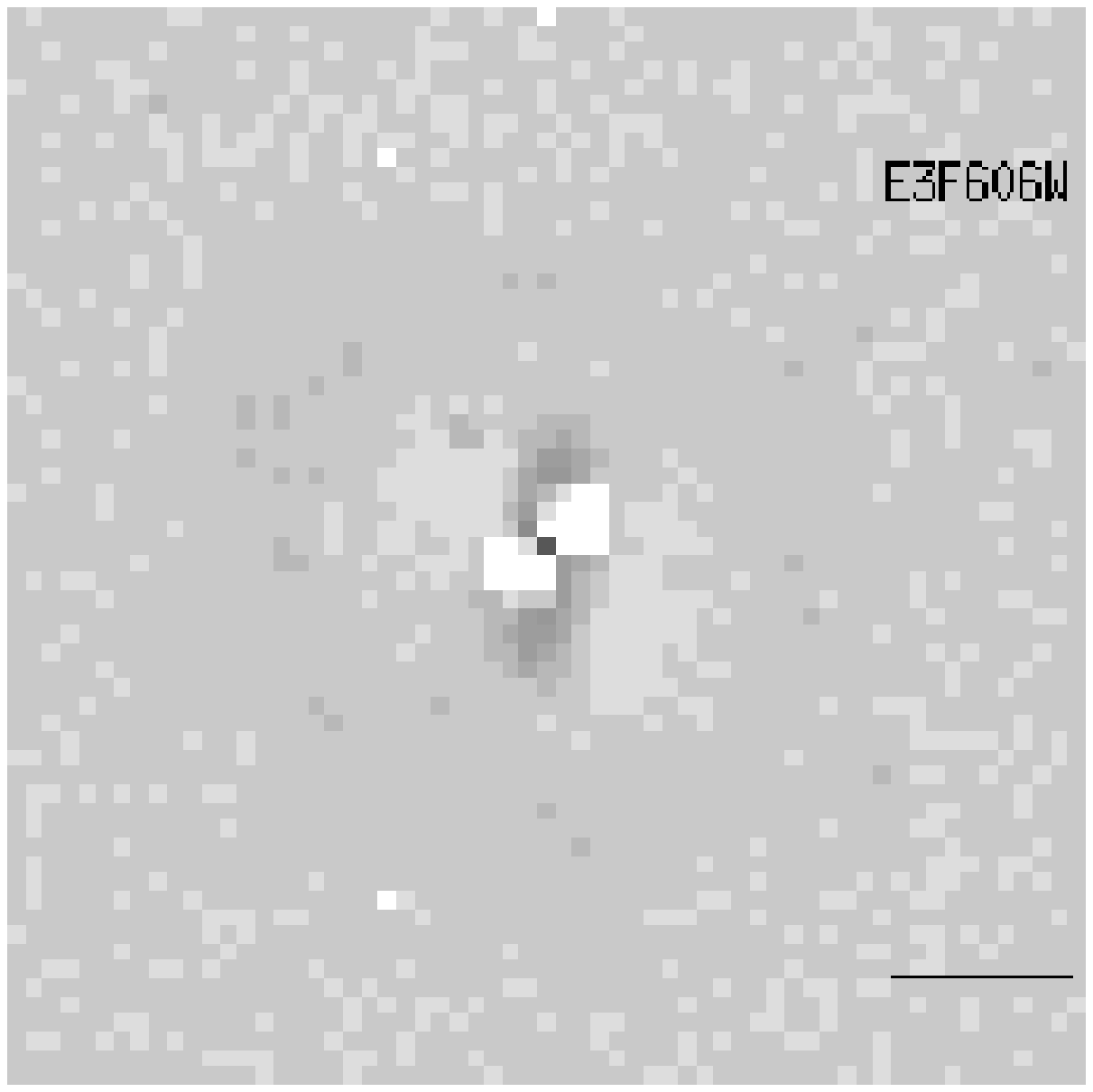}}
\mbox{\epsfysize=4cm \epsfbox{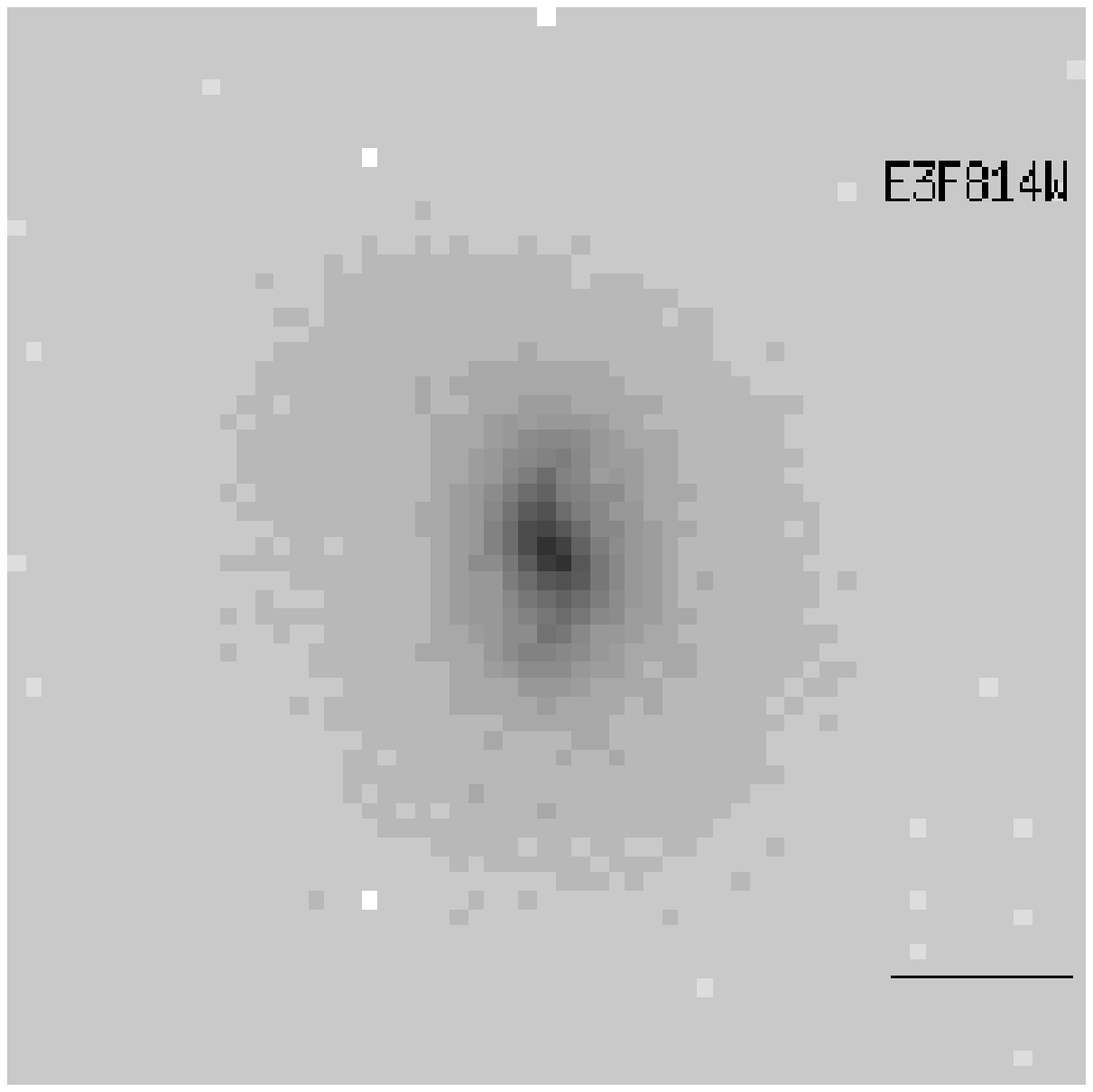}}
\mbox{\epsfysize=4cm \epsfbox{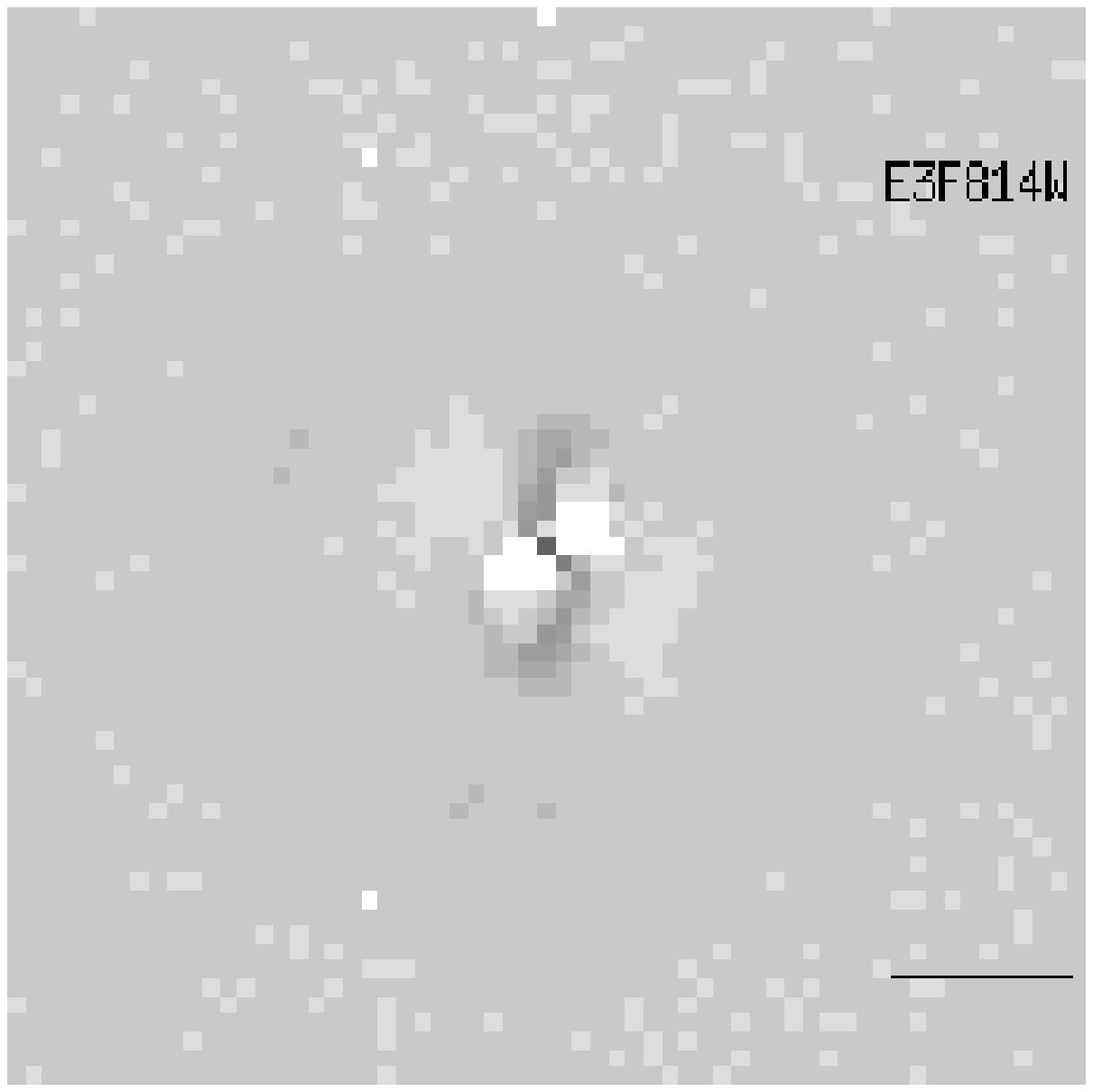}}}
\caption{WFPC2 images and 2D fit residuals of secondary target {\bf
F3}. The horizontal bar is 1 arcsec long. }
\label{fig:2D2}
\end{figure*}

\begin{figure*}
\mbox{
\mbox{\epsfysize=4cm \epsfbox{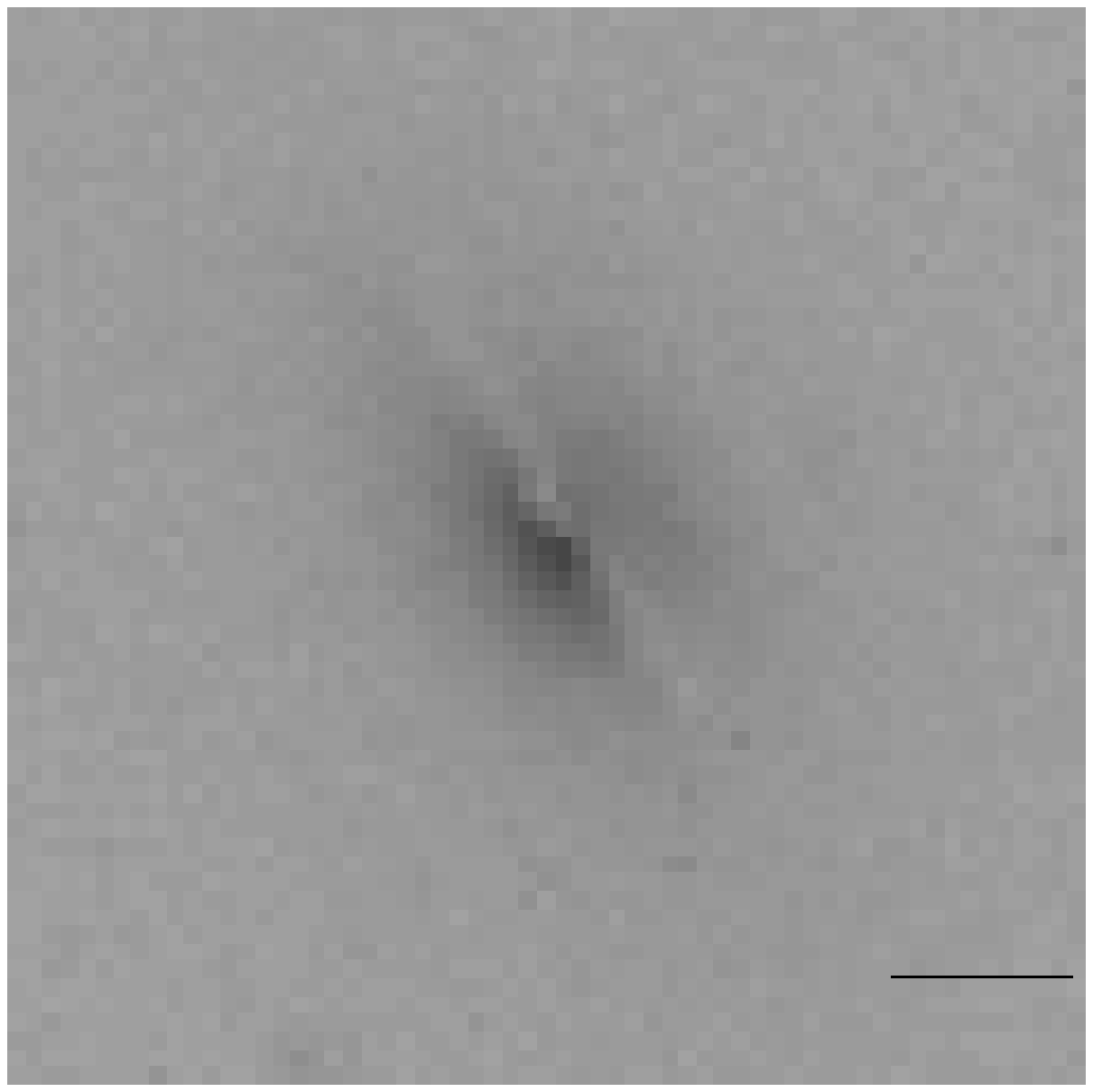}}
\mbox{\epsfysize=4cm \epsfbox{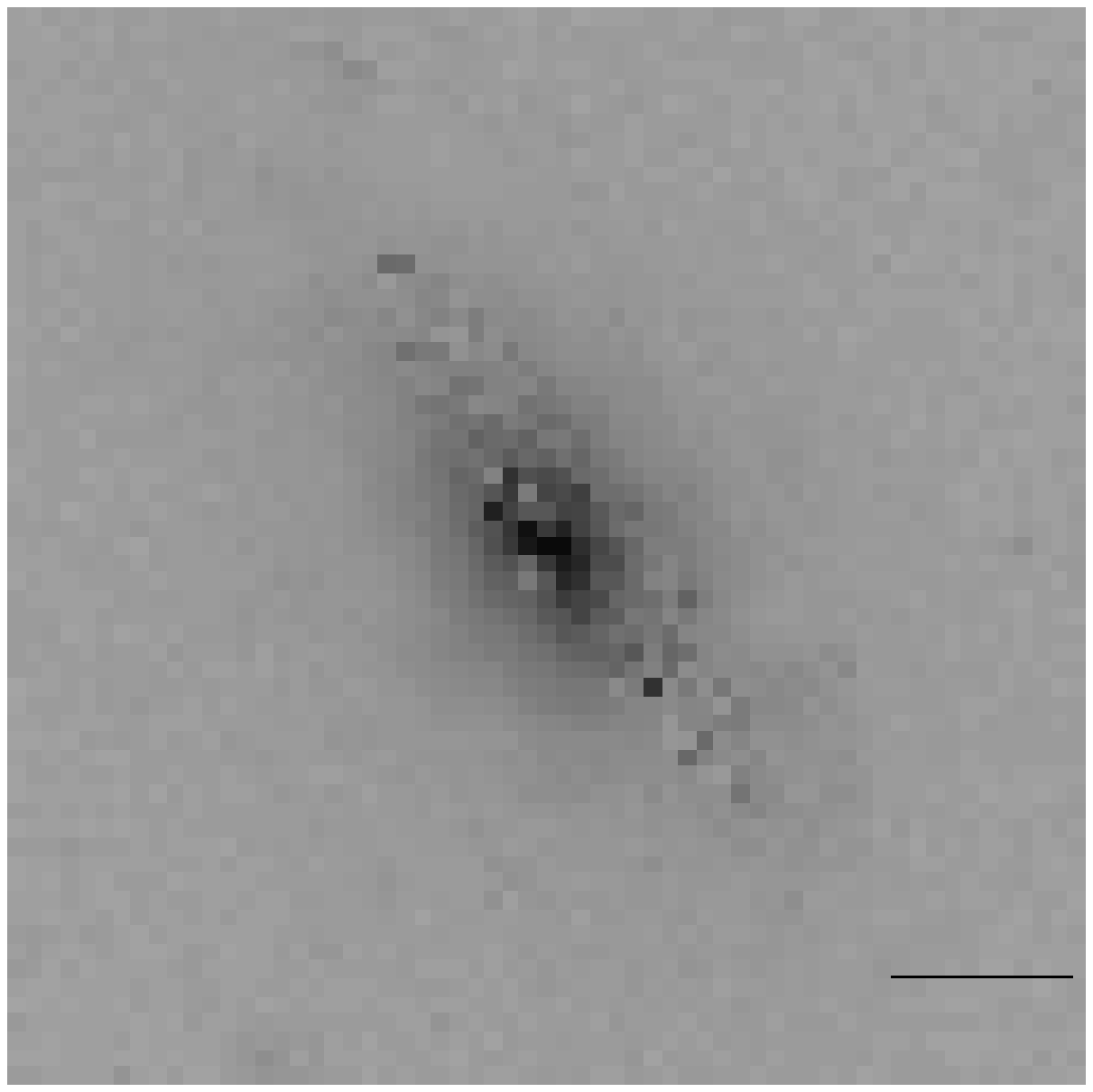}}
\mbox{\epsfysize=4cm \epsfbox{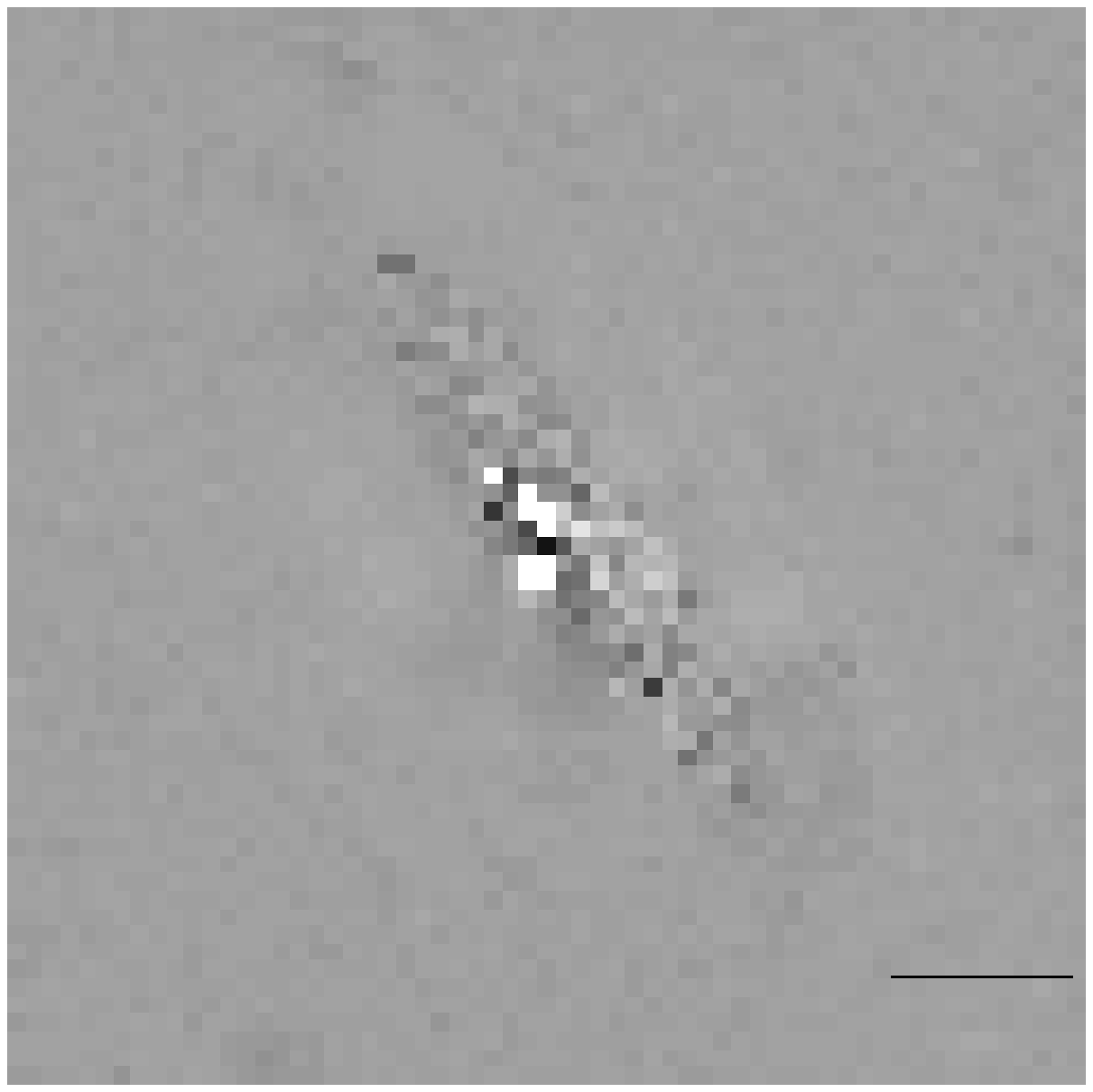}}}
\mbox{
\mbox{\epsfysize=4cm \epsfbox{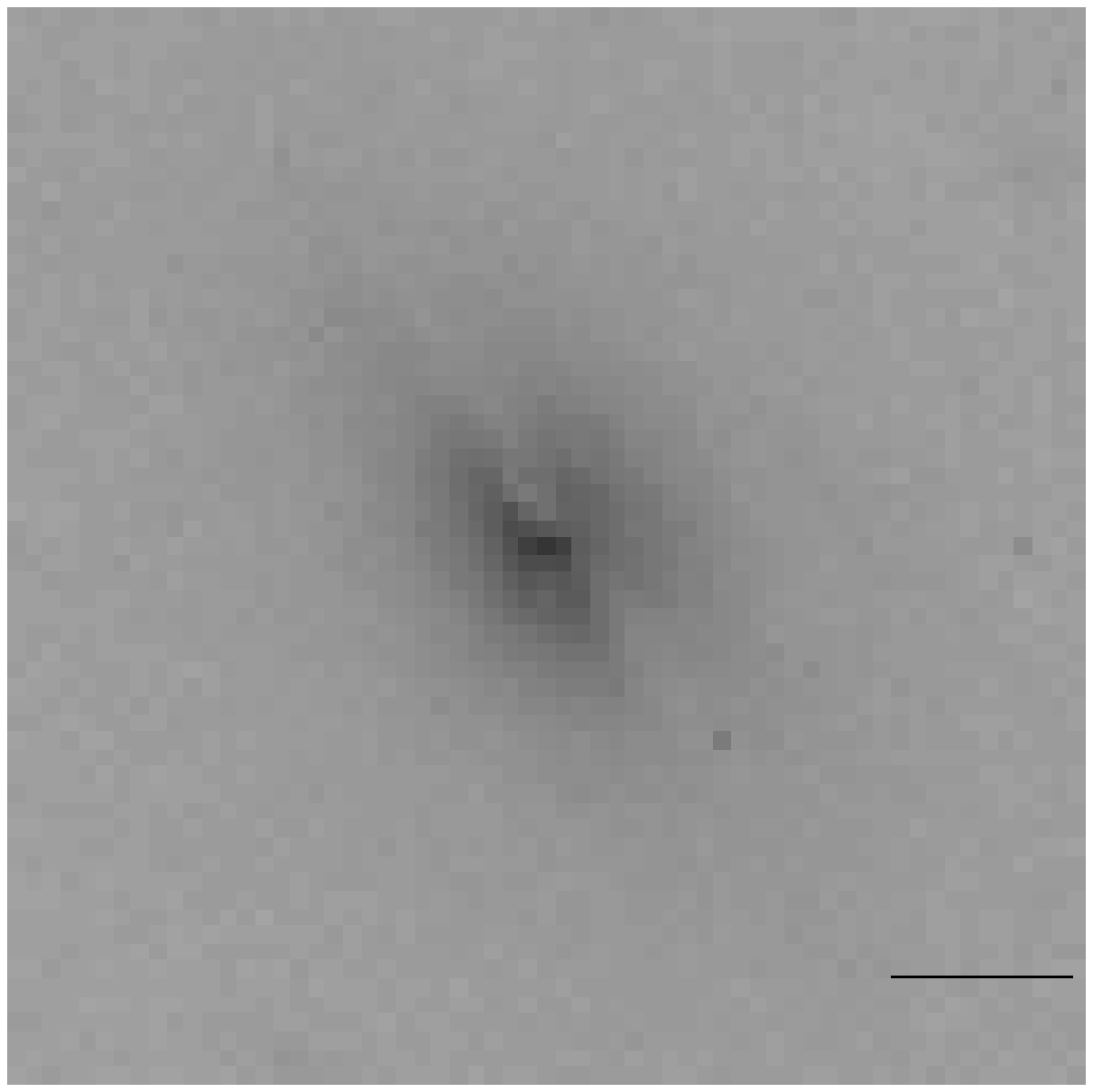}}
\mbox{\epsfysize=4cm \epsfbox{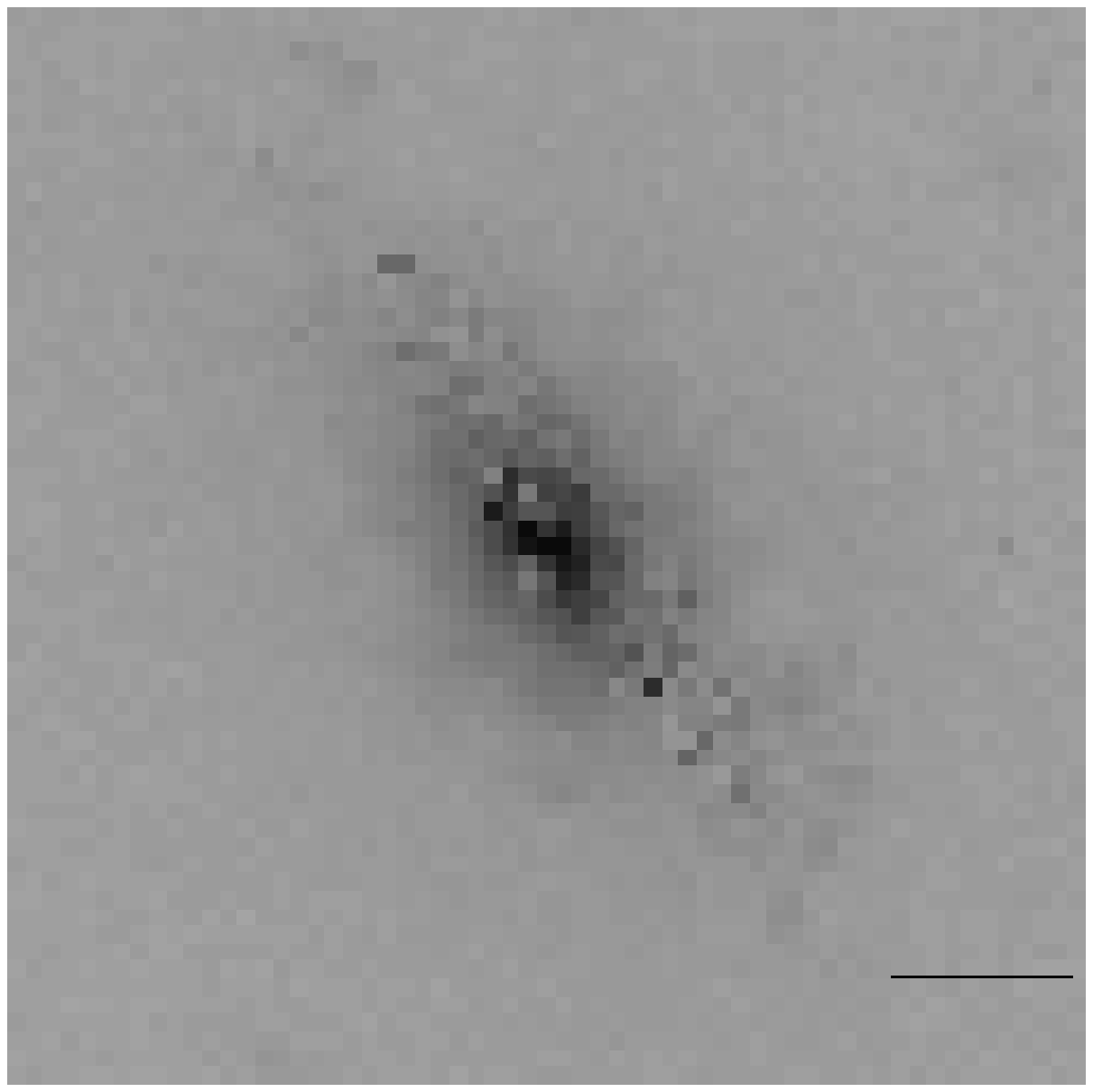}}
\mbox{\epsfysize=4cm \epsfbox{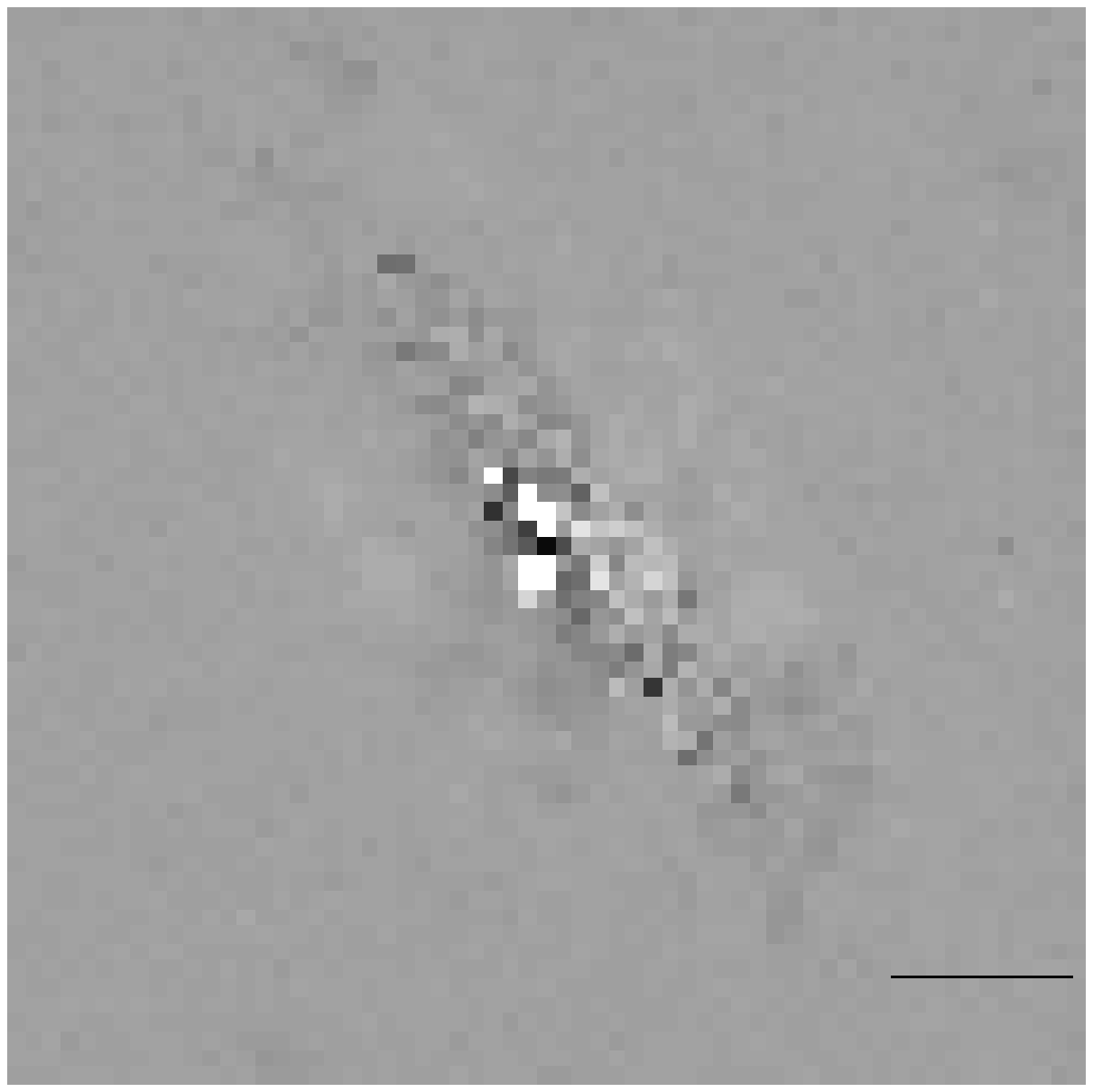}}}
\label{fig:F3}
\caption{WFPC2 images and 2D fit residuals of galaxy {\bf F3}. The
first line shows the F606W images, the image corrected for dust
extinction (see Section~\ref{sec:photo}), and the residuals from the
2D fit. The same images are shown for filter F814W on the second line.
The horizontal bar is 1 arcsec long. The noise in the central part of
the images on the second and third column is higher than the rest,
because of the additional noise due to the internal extinction
correction.}
\end{figure*}

\clearpage

\end{document}